\documentclass[aps,twocolumn,showpacs,preprintnumbers,floats]{revtex4}\def\@cite#1#2{\textsuperscript{[{#1\if@tempswa , #2\fi}]}}

\usepackage{graphicx}
\usepackage{amsmath}
\usepackage{amsfonts}
\usepackage{amssymb}
\usepackage{color}
\usepackage{subfigure}
\usepackage{epsfig}
\usepackage{morefloats}
\usepackage{multirow}
\usepackage{graphicx,booktabs}
\usepackage{mathrsfs}
\usepackage{txfonts}
\usepackage{indentfirst}
\usepackage{graphicx,booktabs}

\usepackage{longtable,lscape}
\usepackage[colorlinks, citecolor=blue,anchorcolor=red,menucolor=red, linkcolor=red,filecolor=red,urlcolor=blue,frenchlinks=red]{hyperref}

\newcommand{\vsig}{\mbox{\boldmath$\sigma$\unboldmath}}

\newcommand{\vrho}{\mbox{\boldmath$\rho$\unboldmath}}
\newcommand{\vlab}{\mbox{\boldmath$\lambda$\unboldmath}}
\newcommand{\vell}{\mbox{\boldmath$\ell$\unboldmath}}

\usepackage{comment}

\begin{document}

%\begin{spacing}{2.0}

\title{Toward discovering  low-lying $P$-wave excited $\Sigma_c$ baryon states}

\author{Kai-Lei Wang$^{1,3}$~\footnote {wangkaileicz@foxmail.com},  and Xian-Hui Zhong$^{2,3}$~\footnote {zhongxh@hunnu.edu.cn}}

\affiliation{ 1) Department
of Electronic Information and Physics, Changzhi University, Changzhi, Shanxi,046011,China}

\affiliation{ 2) Department of
Physics, Hunan Normal University, and Key Laboratory of
Low-Dimensional Quantum Structures and Quantum Control of Ministry
of Education, Changsha 410081, China }
\affiliation{ 3) Synergetic
Innovation Center for Quantum Effects and Applications (SICQEA),
Hunan Normal University, Changsha 410081, China}

%\date{\today}

\begin{abstract}

In this study, by combining the equal spacing rule with  recent observations of
$\Omega_c(X)$ and $\Xi_c(X)$ baryons, we predict the spectrum
of the low-lying $\lambda$-mode $1P$-wave excited $\Sigma_c$ states. Furthermore, their strong decay properties
are predicted using the chiral quark model and the nature of $\Sigma_c(2800)$ is investigated
by analyzing the $\Lambda_c\pi$ invariant mass spectrum. The
$\Sigma_c(2800)$ structure observed in the $\Lambda_c \pi$ mass
spectrum was found to potentially arise from two overlapping $P$-wave $\Sigma_c$ resonances,
$\Sigma_c(2813)3/2^-$ and $\Sigma_c(2840)5/2^-$.
These  resonances have similar decay widths of $\Gamma\sim 40$ MeV
and predominantly decay into the $\Lambda_c \pi$ channel.
The $\Sigma_c(2755)1/2^-$ state is likely to be a very narrow state with
a width of $\Gamma\sim 15$ MeV, with its decays almost saturated by
the $\Lambda_c \pi$ channel. Additionally,  evidence of the $\Sigma_c(2755)\frac{1}{2}^-$ resonance as a very narrow
peak may be seen in the $\Lambda_c\pi$ invariant mass spectrum. The other two $P$-wave states, $\Sigma_c(2746)\frac{1}{2}^-$
and $\Sigma_c(2796)\frac{3}{2}^-$, are relatively narrow states with similar widths
of $\Gamma\sim 30$ MeV and predominantly decay into $\Sigma_c\pi$ and $\Sigma^{*}_c\pi$, respectively.
We hope our study can provide useful references for discovering these low-lying $P$-wave states
in forthcoming experiments.

\end{abstract}


\maketitle

\section{Introduction}

Over the past several years, immense progress toward the observations of singly heavy baryons has
been achieved at the LHC.  In 2017, five extremely narrow $\Omega_c(X)$
states, $\Omega_c(3000)$, $\Omega_c(3050)$, $\Omega_c(3066)$, $\Omega_c(3090)$, and $\Omega_c(3119)$,
were observed in the $\Xi_c^{+}K^-$ channel by the LHCb Collaboration~\cite{Aaij:2017nav}.
In 2018, the LHCb Collaboration observed a new bottom baryon, $\Xi_b(6227)^-$, in both
$\Lambda_b^0K^-$ and $\Xi_b^0\pi^-$ decay modes~\cite{Aaij:2018yqz} and two new resonances, $\Sigma_b(6097)^{\pm}$,
in the $\Lambda_b^0 \pi^{\pm}$ channels~\cite{Aaij:2018tnn}.
In 2020, the LHCb Collaboration also observed four new $\Omega_b(X)$ states,
$\Omega_b(6316)^-$, $\Omega_b(6330)^-$, $\Omega_b(6340)^-$, and $\Omega_b(6350)^-$, in
the $\Xi^0_bK^-$ mass spectrum~\cite{Aaij:2020cex};
three new $\Xi_c(X)$ states, $\Xi_c(2923)^0$, $\Xi_c(2939)^0$, and $\Xi_c(2965)^0$, in the
$\Lambda_c^+K^-$ mass spectrum~\cite{Aaij:2020yyt}; and a new $\Xi_b(6227)^0$ state
in the $\Xi_b^-\pi^+$ channel~\cite{Aaij:2020fxj}. Also in 2020, the CMS Collaboration  observed
a broad enhancement around 6070 MeV in the $\Lambda^0_b\pi^+\pi^-$ invariant mass spectrum~\cite{Sirunyan:2020gtz}, which was confirmed
by  subsequent LHCb experiments with high statistical significance~\cite{Aaij:2020rkw}.
More recently, the CMS Collaboration observed a new excited beauty strange baryon, $\Xi_b(6100)^-$,
decaying to $\Xi^-_\mathrm{b} \pi^+ \pi^-$~\cite{Sirunyan:2021vxz}.

These newly observed resonances provide opportunities for
establishing an abundant singly heavy baryon spectrum.
For a singly heavy baryon, there are two kinds of excitations, ``$\rho$-mode" and ``$\lambda$-mode". The $\rho$-mode excitation appears within the light diquark, while the
$\lambda$-mode excitation occurs between the light diquark and the heavy quark.
In the heavy quark limit,  $m_{Q} \to \infty$,  no mixing occurs between the $\lambda$- and $\rho$-mode excitations due to a
strong suppression of the spin-dependent interactions by the heavy quark mass $m_{Q}$~\cite{Yoshida:2015tia,Capstick:1986bm,Roberts:2007ni}.
The $\rho$-mode excitation energy should be notably larger than that of the $\lambda$-mode~\cite{Liu:2012sj,Bijker:2020tns,Yoshida:2015tia,Chen:2021eyk,Xiao:2020gjo,Wang:2017kfr}, which can be understood using the simple harmonic oscillator model, as the frequency $\omega_{\rho}$ of the $\rho$-mode
is larger than the frequency $\omega_{\lambda}$ of the $\lambda$-mode.
The lower $\lambda$-mode excitation energy indicates that the $\lambda$-mode excitations
should be more easily formed than $\rho$-mode excitations.

In the literature, the masses for the $1P$-wave $\lambda$-mode $\Omega_{c}$, $\Omega_{b}$, $\Xi_{c}'$,  $\Xi_{b}'$,
and $\Sigma_{b}$ baryon states are predicted to be $\sim 2.95-3.10$ GeV~\cite{Karliner:2017kfm,Aliev:2017led,Wang:2017zjw,Perez-Rubio:2015zqb,Chen:2015kpa,Wang:2010it,Wang:2017vnc,Chen:2017gnu,Yoshida:2015tia,
Ebert:2007nw,Ebert:2011kk,Roberts:2007ni,Yamaguchi:2014era,
Shah:2016nxi,Garcilazo:2007eh,Bahtiyar:2020uuj,Padmanath:2017lng}, $\sim 6.30-6.38$ GeV~\cite{Mutuk:2020rzm,Mao:2015gya,Xu:2020ofp,Wang:2010it,Yoshida:2015tia,Ebert:2007nw,Ebert:2011kk,Roberts:2007ni,
Yamaguchi:2014era,Garcilazo:2007eh,Karliner:2020fqe,Wang:2020pri}, $\sim 2.85-3.03$ GeV~\cite{Perez-Rubio:2015zqb,Chen:2015kpa,Bijker:2020tns,Ebert:2007nw,Ebert:2011kk,Roberts:2007ni,Yamaguchi:2014era,Shah:2016nxi,Garcilazo:2007eh,Bahtiyar:2020uuj}, $\sim 6.15-6.25$ GeV~\cite{Mao:2015gya,Bijker:2020tns,Ebert:2007nw,Ebert:2011kk,Roberts:2007ni,Yamaguchi:2014era,Garcilazo:2007eh}, and
$\sim 6.07-6.20$ GeV~\cite{Mao:2015gya,Yoshida:2015tia,Ebert:2007nw,Ebert:2011kk,Roberts:2007ni,
Yamaguchi:2014era,Garcilazo:2007eh,Capstick:1986bm}, respectively. The $1P$-wave $\rho$-mode
states lie $70-150$ MeV above the $\lambda$-mode states according to the
quark model predictions in Refs.~\cite{Liu:2012sj,Bijker:2020tns,Yoshida:2015tia,Chen:2021eyk,Capstick:1986bm}.
It should be mentioned that the $\rho$-mode excitation energy calculated within QCD sum rules is slightly lower
than that of the $\lambda$-mode in some other  cases~\cite{Yang:2019cvw,Chen:2017sci,Yang:2020zrh,Mao:2015gya,Chen:2015kpa,Cui:2019dzj,Yang:2020zjl,Chen:2020mpy}.
The newly observed singly heavy baryons, $\Omega_c(X)$, $\Omega_b(X)$, $\Xi_c(X)$, $\Xi_b(6227)^0$, and $\Sigma_b(6097)^{\pm}$, are just within the predicted mass ranges of the $1P$-wave excitations. Furthermore, prompted by the newly observed singly heavy baryon
states and combined with mass spectrum,  the strong decay properties have been studied using
the QCD sum rules~\cite{Aliev:2018vye,Yang:2019cvw,Agaev:2017lip,Chen:2017sci,Cui:2019dzj,Yang:2020zjl,Chen:2020mpy,Yang:2020zrh}, $^3P_0$ model~\cite{Santopinto:2018ljf,Bijker:2020tns,Chen:2017gnu,Chen:2018vuc,Chen:2018orb,Yang:2018lzg,Lu:2020ivo,He:2021xrh,Liang:2020hbo}, chiral quark model~\cite{Wang:2020gkn,Xiao:2020oif,Wang:2018fjm,Xiao:2020gjo,Wang:2017kfr,Wang:2017hej}, and heavy quark effective
theory~\cite{Cheng:2017ove,Wang:2017vnc}.

Based on the mass spectrum and strong decay analyses, the $\Omega_c(3000)$, $\Omega_c(3050)$, $\Omega_c(3066)$, $\Omega_c(3090)$, and
$\Omega_c(3119)$ structures may be explained with the $1P$-wave $\lambda$-mode $\Omega_{c}$ states~\cite{Karliner:2017kfm,Wang:2017zjw,Padmanath:2017lng,Chen:2017gnu,Wang:2017vnc},
although there are different explanations about some of the resonances, such as $\Omega_c(3090)$ and $\Omega_c(3119)$,
which may be explained with  radially excited ($2S$-wave) $\Omega_c$ states~\cite{Wang:2017hej,Cheng:2017ove,Agaev:2017jyt,Agaev:2017lip}.
It should be mentioned that the recent LHCb measurements show that the spin assignment
of the four observed states $\Omega_c(3000)$, $\Omega_c(3050)$, $\Omega_c(3066)$, and $\Omega_c(3090)$
is consistent with $\lambda$-mode excitations with quantum numbers $J=1/2, 3/2, 3/2$ and 5/2~\cite{LHCb:2021ptx}.
It is interesting to notice that among various models, only the predicted $J^P$ quantum numbers
in our previous work~\cite{Wang:2017hej} are consistent with the above-mentioned
scenario as pointed in the recent review of charmed  baryon physics~\cite{Cheng:2021qpd}. Similarly, the new $\Omega_b(X)$ states, $\Omega_b(6316)^-$, $\Omega_b(6330)^-$, $\Omega_b(6340)^-$, and $\Omega_b(6350)^-$, can be assigned to the $1P$-wave $\lambda$-mode $\Omega_{b}$ states~\cite{Wang:2020pri,Karliner:2020fqe,Mutuk:2020rzm,Liang:2020hbo}, although the $\Omega_b(6316)^-$ may
be a $\rho$-mode excitation, as suggested in~\cite{Yang:2020zrh}. The new $\Xi_c(X)$ states,
$\Xi_c(2923)^0$, $\Xi_c(2939)^0$, and $\Xi_c(2965)^0$, are also good candidates
for the $1P$-wave $\lambda$-mode $\Xi_{c}'$ states belonging to $\mathbf{6}_F$, as suggested
in the literature~\cite{Wang:2020gkn,Yang:2020zjl,Bijker:2020tns},
although different explanations exist for some resonances,
such as $\Xi_c(2939)^0$ and $\Xi_c(2965)^0$, which  may be candidates of the $1P$-wave $\rho$-mode
excitations~\cite{Bijker:2020tns}, and $\Xi_c(2965)^0$, which may be the $J^P=1/2^+$ $\Xi_{c}'(2S)$
state~\cite{Lu:2020ivo,Agaev:2020fut}. Additionally, the $\Sigma_b(6097)^{\pm}$ and
$\Xi_b(6227)^0$ resonances are good candidates for the $1P$-wave $\lambda$-mode
singly bottom baryons~\cite{Jia:2019bkr,Wang:2018fjm,Yang:2019cvw,Yang:2020zrh,Xiao:2020gjo,
He:2021xrh,Cui:2019dzj,Aliev:2018vye,Yang:2018lzg,Chen:2018orb,Chen:2018vuc}.
Finally,
some unconventional interpretations, such as molecular
or pentaquark, were also proposed in the literature for the newly observed resonances, $\Omega_c(X)$~\cite{Wang:2018alb,Huang:2017dwn,Kim:2017jpx,An:2017lwg,Kim:2017khv,Montana:2017kjw,Debastiani:2017ewu,Huang:2018wgr,
Debastiani:2018adr,Liu:2018bkx,Montana:2018edp,Wang:2021cku}, $\Omega_b(X)$~\cite{Liang:2020dxr},
$\Xi_c(X)$~\cite{Hu:2020zwc,Zhu:2020jke}, and $\Xi_b(6227)^-$~\cite{Huang:2018bed,Yu:2018yxl,Wang:2020vwl,Zhu:2020lza}.
As a whole, a fairly complete $\lambda$-mode $P$-wave spectrum in the $\Omega_c$, $\Xi_c'$, $\Xi'_b$, $\Omega_b$,
and $\Sigma_b$ families may  be established with discovery of the series of heavy baryons at the LHC.
Based on our previous work~\cite{Xiao:2020gjo,Wang:2020gkn,Xiao:2020oif,Wang:2018fjm,Wang:2017kfr,Wang:2017hej},
we provide a quark model classification of these newly observed resonances,
summarized in Table~\ref{classfy}.

\begin{table*}[ht]
\caption{Quark model classifications of the newly observed singly heavy baryon resonances based on our previous work~\cite{Wang:2020gkn,Xiao:2020oif,Wang:2018fjm,Wang:2017kfr,Wang:2017hej}. This table is taken from~\cite{Xiao:2020gjo}.} \label{classfy}
\begin{tabular}{cccccccccccccccc }\hline\hline
$L$-$S$ scheme& $j$-$j$ scheme & \multicolumn{6}{c}{Observed states/structures belonging to the $\mathbf{6}_F$ multiplet.} \\ \hline
~~~~$|n^{2S+1}L_{\lambda}J^P\rangle$~~~~~~& ~~~~$|J^P,j\rangle~(nl)$  ~~~~~~~~~~~~&$\Omega_c$ states~~~~~~&$\Xi_c'$ states~~~~~~&$\Sigma_c$ states
~~~~~~&$\Omega_b$ states~~~~~~&$\Xi_b'$ states~~~~~~&$\Sigma_b$ states \\
\hline
$|1^{2}S \frac{1}{2}^+\rangle$ &$|J^P=\frac{1}{2}^+,1\rangle(1S)$~~&$\Omega_c(2695)$ ~~&$\Xi_c'(2578)$~~&$\Sigma_c(2455)$~~&$\Omega_b(6046)$~~&$\Xi_b'(5935)$~~&$\Sigma_b(5810)$\\
$|1^{4}S \frac{3}{2}^+\rangle$ &$|J^P=\frac{3}{2}^+,1\rangle(1S)$~~&$\Omega_c^*(2770)$ ~~&$\Xi_c^*(2645)$~~&$\Sigma_c^*(2520)$~~&$\cdots$~~&$\Xi_b^*(5955)$~~&$\Sigma_b^*(5830)$\\
$|1P_\lambda \frac{1}{2}^-\rangle_1$ &$|J^P=\frac{1}{2}^-,1\rangle(1P)$~~&$\Omega_c(3000)$ ~~&$\cdots$~~&$\cdots$~~&$\Omega_b(6316)$~~&$\cdots$~~&$\Sigma_b(6072)$  \\
$|1P_\lambda \frac{1}{2}^-\rangle_2$ &$|J^P=\frac{1}{2}^-,0\rangle(1P)$~~&$\cdots$ ~~&$\Xi_c(2880)$~~&$\cdots$~~&$\cdots$~~&$\cdots$~~&$\cdots$~~\\
$|1^{4}P_\lambda \frac{3}{2}^-\rangle$ &$|J^P=\frac{3}{2}^-,1\rangle(1P)$~~&$\Omega_c(3050)$ ~~&$\Xi_c(2923)$~~&$\cdots$~~&$\Omega_b(6330)$~~&$\cdots$~~&$\Sigma_b(6072)$ \\
$|1^{2}P_\lambda \frac{3}{2}^-\rangle$ &$|J^P=\frac{3}{2}^-,2\rangle(1P)$~~&$\Omega_c(3065)$ ~~&$\Xi_c(2939)$~~&$\Sigma_c(2800)$~~&$\Omega_b(6340)$~~&$\Xi_b'(6227)$~~&$\Sigma_b(6097)$ \\
$|1^{4}P_\lambda \frac{5}{2}^-\rangle$ &$|J^P=\frac{5}{2}^-,2\rangle(1P)$~~&$\Omega_c(3090)$ ~~&$\Xi_c(2965)$~~&$\Sigma_c(2800)$~~&$\Omega_b(6350)$~~&$\Xi_b'(6227)$~~&$\Sigma_b(6097)$ \\
\hline\hline
\end{tabular}
\end{table*}

LHC experiments have demonstrated the capability for the discovery of heavy baryons. Therefore, the missing $\lambda$-mode $P$-wave $\Sigma_c$ baryon states
are  likely to be discovered by forthcoming LHC experiments.
The $\Sigma_c$ mass spectrum has been studied theoretically using various approaches, such as the relativized quark model~\cite{Capstick:1986bm}, relativistic quark model~\cite{Ebert:2007nw,Ebert:2011kk,Migura:2006ep}, non-relativistic quark model~\cite{Roberts:2007ni,Chen:2016iyi,Shah:2016mig,Shah:2016nxi,Yoshida:2015tia,Jia:2019bkr}, lattice QCD~\cite{Bahtiyar:2020uuj,Perez-Rubio:2015zqb},
QCD sum rules~\cite{Wang:2010it,Zhang:2008iz,Chen:2015kpa}, and more.
Some quark model predictions of the masses for the $\lambda$-mode $S$-wave and $P$-wave $\Sigma_c$ states are collected in Table~\ref{sp1}~\cite{Chen:2016iyi,Ebert:2011kk,Shah:2016mig,Yamaguchi:2014era,Ebert:2007nw,Capstick:1986bm}.
Using the heavy-quark-light-diquark approximation, the masses of the $\lambda$-mode $P$-wave $\Sigma_c$ states in the relativistic quark model
are predicted to be approximately  $2.71-2.81$ GeV~\cite{Ebert:2007nw,Ebert:2011kk}, which is consistent with that of the non-relativistic quark model~\cite{Chen:2016iyi}. With the hypercentral approximation, the $\lambda$-mode $P$-wave $\Sigma_c$ states in the non-relativistic quark model
are predicted to be approximately  $2.79-2.84$ GeV~\cite{Shah:2016mig}.
By strictly solving the three body problem without the diquark and hypercentral approximations, the masses of the $\lambda$-mode $P$-wave $\Sigma_c$ states are predicted to be approximately  $2.76-2.82$ GeV and $2.80-2.84$ GeV in the relativized quark model~\cite{Capstick:1986bm} and non-relativistic quark model~\cite{Yoshida:2015tia}, respectively. The masses for the two $\rho$-mode $P$-wave $\Sigma_c$ states
with $J^P=1/2^-$ and $3/2^-$ are predicted to be $\sim 2.85-2.91$ GeV~\cite{Yoshida:2015tia,Capstick:1986bm},
which are approximately $70$ MeV larger than the highest $\lambda$-mode excitation.
Considering the mass, the $\Sigma_c(2800)$ resonance~\cite{Tanabashi:2018oca} observed in the $\Lambda_c \pi$ final states by the Belle and $BABAR$ Collaborations~\cite{Mizuk:2004yu,Aubert:2008ax} may be  experimental signals of
the $P$-wave $\Sigma_c$ states. The case of $\Sigma_c(2800)$ as the $\rho$-mode
$P$-wave excitations should be excluded as the $\Lambda_c \pi$ decay channel is
forbidden~\cite{Zhong:2007gp}. There are some discussions on the nature of $\Sigma_c(2800)$ in the
literature~\cite{Wang:2017kfr,Zhong:2007gp,Cheng:2015iom,Cheng:2006dk,Chen:2007xf,Ebert:2007nw,
Gerasyuta:2007un,Garcilazo:2007eh,Chen:2015kpa,Chen:2017sci,Cheng:2021qpd}, however, these involve strong model dependencies.
For example, the spin-parity ($J^P$) numbers were suggested to be $J^P=3/2^-$ within the heavy hadron chiral
perturbation theory approach~\cite{Cheng:2006dk,Cheng:2021qpd}, $J^P=3/2^-$ or $J^P=5/2^-$ in the $^3P_0$ model~\cite{Chen:2007xf},
and $J^P=1/2^-$ or $3/2^-$ using the QCD sum rule approach~\cite{Chen:2015kpa,Chen:2017sci}. In our previous study,
it was found that $\Sigma_c(2800)$ might favor the $J^P=3/2^-$ state
$|\Sigma_c~ ^2P_{\lambda} \frac{3}{2}^- \rangle$ or the $J^P=5/2^-$ state
$|\Sigma_c~ ^4P_{\lambda} \frac{5}{2}^- \rangle$ in the $L$-$S$ coupling scheme~\cite{Wang:2017kfr}.
%It should be pointed out that the $\Sigma_c(2800)$ structure observed in the $\Lambda_c \pi$ mass
%spectrum might arise from several overlapping $P$-wave $\Sigma_c$ resonances with different
%spin-parity numbers.

In this study, we revisit the $\lambda$-mode $P$-wave $\Sigma_c$ baryon states. The main aims are as follows:
(i) the spectrum was classified in the $L$-$S$ coupling scheme in
our previous work~\cite{Wang:2017kfr}, where configuration mixing between two different
states with the same $J^P$ numbers, which may be caused by antisymmetric spin-orbit forces, is not considered.
This configuration mixing may notably affect some of our predictions,
thus, we include this effect here by adopting the $j$-$j$ coupling scheme.
(ii) We hope to provide more reliable predictions for
the $\lambda$-mode $P$-wave $\Sigma_c$ baryon states by combining the information from
the most recent observations of the $\Xi_c(X)$ and $\Omega_c(X)$ states.

This paper is organized as follows. In Sec.~\ref{MAL}, we provide a quark model classification of the
singly heavy baryon states and the mass analysis of the $\lambda$-mode $1P$-wave $\Sigma_c$
states by incorporating the recent observations of the singly-heavy baryons.
Then, according to our chiral quark model calculations,  their strong decay properties are discussed in Sec.~\ref{STD}.
To determine the contributions of the $P$-wave $\Sigma_c$ states
to the experimentally observed $\Sigma_c(2800)$ resonance~\cite{Tanabashi:2018oca}, we further analyze the
$\Lambda_c\pi$ invariant mass spectrum measured by $BABAR$~\cite{Aubert:2008ax} in Sec.~\ref{IVR}.
Finally, we summarize our results in Sec.~\ref{SUM}.

\begin{table*}[htp]
\begin{center}
\caption{\label{sp1}   Predicted mass spectrum of $1S$-wave and $\lambda$-mode $1P$-wave $\Sigma_c$ states belonging to the $\mathbf{6}_F$ multiplet in various quark models. The $\Sigma_c$ states are denoted by $|J^P,j\rangle$ in the $j$-$j$ coupling scheme,
where $j$ stands for the total angular momentum quantum number of the two light quarks. The unit of mass is MeV.}
\begin{tabular}{lccccccccccccccccccccccccccccccccccccccccccccc}\hline\hline
%State                            ~~~~ &             ~~~~ &Predicted                ~~~~&Predicted  ~~~~&Predicted   ~~~~&Predicted  ~~~~&Predicted  ~~~~&Predicted ~~~~&        \\
~~~~$|J^P,j\rangle$ ~~~~&Ref.~\cite{Ebert:2007nw}   ~~~~&Ref.~\cite{Ebert:2011kk}  ~~~~
&Ref.~\cite{Yoshida:2015tia}~~~~&Ref.~\cite{Capstick:1986bm}~~~~&Ref.~\cite{Chen:2016iyi} ~~~~&Ref.~\cite{Shah:2016mig}~~~~
  ~~~~&Observed state \\ \hline
$|J^P=\frac{1}{2}^+,1\rangle$   ~~~~&2439~~~~&2443~~~~&2460             ~~~~&2440 ~~~~&2456~~~~&2452~~~~ &$\Sigma_c(2455)$ \\
$|J^P=\frac{3}{2}^+,1\rangle$   ~~~~&2518~~~~&2519~~~~&2523             ~~~~&2495 ~~~~&2515~~~~&2501~~~~ &$\Sigma^*_c(2520)$  \\
$|J^P=\frac{1}{2}^-,0\rangle$   ~~~~&2795~~~~&2713~~~~&2802             ~~~~&2765 ~~~~&2702~~~~&2832~~~~ &$\Sigma_c(2755)$ ? \\
$|J^P=\frac{1}{2}^-,1\rangle$   ~~~~&2805~~~~&2799~~~~&2826             ~~~~&2770 ~~~~&2765~~~~&2841~~~~ &$\Sigma_c(2746)$ ?  \\
$|J^P=\frac{3}{2}^-,1\rangle$   ~~~~&2761~~~~&2773~~~~&2807             ~~~~&2770 ~~~~&2785~~~~&2812~~~~ &$\Sigma_c(2796)$ ?\\
$|J^P=\frac{3}{2}^-,2\rangle$   ~~~~&2799~~~~&2798~~~~&2837             ~~~~&2805 ~~~~&2798~~~~&2822~~~~ &$\Sigma_c(2813)$ ? \\
$|J^P=\frac{5}{2}^-,2\rangle$   ~~~~&2790~~~~&2789~~~~&2839             ~~~~&2815 ~~~~&2790~~~~&2796~~~~ &$\Sigma_c(2840)$ ?   \\
\hline\hline
\end{tabular}
\end{center}
\end{table*}

\begin{table}[htp]
\begin{center}
\caption{\label{JJ}   Classification of the $\lambda$-mode $1P$-wave singly heavy baryon states belonging to $\mathbf{6}_F$ in the $j$-$j$ coupling scheme. The states in the $j$-$j$ coupling scheme are denoted by $|J^P,j \rangle$.}
\begin{tabular}{p{2.1cm}p{0.6cm}p{0.6cm}p{0.6cm}p{0.6cm}p{0.6cm}p{0.6cm}p{0.6cm}p{0.6cm}p{0.6cm} |}\hline\hline
~~~~$|J^P,j \rangle$      &$J^P$            &$j$   &$\ell_{\rho}$    &$\ell_{\lambda}$    &$L$   &$s_{\rho}$    &$s_{Q}$                  &$S$    \\ \hline
$|J^P=\frac{1}{2}^-,0\rangle$   &$\frac{1}{2}^-$  & 0        & 0                &1          &1       &1         &$\frac{1}{2}$         &$\frac{1}{2}$,$\frac{3}{2}$\\
$|J^P=\frac{1}{2}^-,1\rangle$   &$\frac{1}{2}^-$  & 1        & 0                &1          &1       &1         &$\frac{1}{2}$         &$\frac{1}{2}$,$\frac{3}{2}$     \\
$|J^P=\frac{3}{2}^-,1\rangle$   &$\frac{3}{2}^-$  & 1        & 0                &1          &1       &1         &$\frac{1}{2}$         &$\frac{1}{2}$,$\frac{3}{2}$  \\
$|J^P=\frac{3}{2}^-,2\rangle$   &$\frac{3}{2}^-$  & 2        & 0                &1          &1       &1         &$\frac{1}{2}$         &$\frac{1}{2}$,$\frac{3}{2}$   \\
$|J^P=\frac{5}{2}^-,2\rangle$   &$\frac{5}{2}^-$  & 2        & 0                &1          &1       &1         &$\frac{1}{2}$         &$\frac{3}{2}$   \\
\hline\hline
\end{tabular}
\end{center}
\end{table}

\begin{figure}[htbp]
\centering \epsfxsize=4.0cm \epsfbox{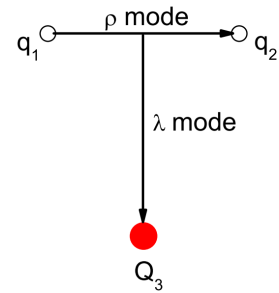}  \vspace{-0.3 cm}\caption{(Color online) $q_1q_2Q_3$ system with $\lambda$- or $\rho$-mode excitations.}\label{mode}
\end{figure}

\begin{figure}[htbp]
\centering \epsfxsize=9.6cm \epsfbox{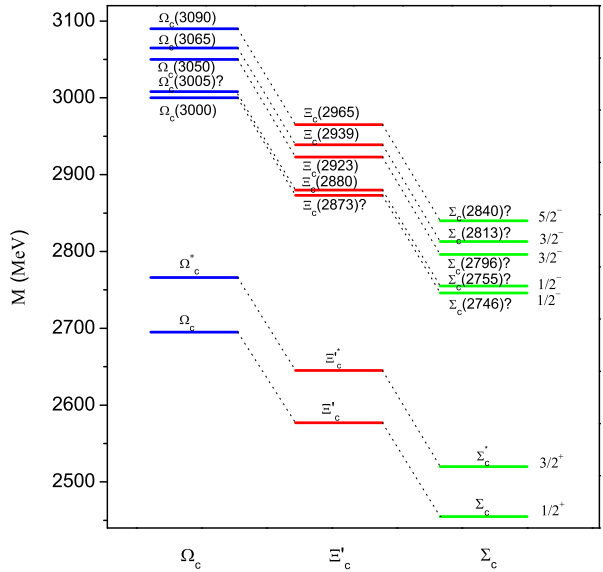}  \vspace{-1.0 cm}\caption{ Mass spectra of the $1S$ and $1P$-wave $\Omega_c$, $\Xi_c'$ and $\Sigma_c$ baryon states belonging to $\mathbf{6}_F$, predicted by combining the equal spacing rule with the recent observations of
the $\Omega_c(X)$ and $\Xi_c(X)$ baryons.}\label{mass}
\end{figure}

\section{Mass spectrum analysis}\label{MAL}

\subsection{Quark model classification}

For a singly heavy baryon system $q_1q_2Q_3$, shown in Fig.\ref{mode},
it is convenient to introduce two Jacobi coordinates,
\begin{eqnarray}
\vrho &=&\frac{1}{\sqrt{2}}(\mathbf{r}_1-\mathbf{r}_2),\\
\vlab &=&\frac{1}{\sqrt{6}}(\mathbf{r}_1+\mathbf{r}_2-2\mathbf{r}_3),
\end{eqnarray}
where $\mathbf{r}_{1}$ and $\mathbf{r}_{2}$ are coordinates for the light quarks $q_1$
and $q_2$, respectively, while $\mathbf{r}_3$ is the coordinate for the heavy quark $Q_3$.
The orbital/radial excitation appearing between the light quarks $q_1$ and
$q_2$ with a Jacobi coordinate $\vrho$ is denoted by ``$\rho$-mode",
while the excitation appearing between the light diquark $q_1q_2$ and
heavy quark $Q_3$ with a Jacobi coordinate $\vlab$ is denoted by
``$\lambda$-mode".

The heavy baryon containing a heavy quark violates
the SU(4) symmetry. However, the SU(3) symmetry between the other two
light quarks ($u$, $d$, or $s$) is approximately kept. According to
this symmetry,  heavy baryons containing a single heavy quark
belong to two different SU(3) flavor representations: the symmetric sextet $\mathbf{6}_F$ and
antisymmetric antitriplet $\bar{\mathbf{3}}_F$~\cite{Wang:2017kfr}.
For charmed baryons $\Lambda_c$ and $\Xi_c$ belonging to $\bar{\mathbf{3}}_F$, the antisymmetric flavor
wave functions can be written as
\begin{equation} \phi^c_{\bar{\mathbf{3}}}=\begin{cases}
        \frac{1}{\sqrt{2}}(ud-du)c   &$for$~\Lambda_c^{+},\\
        \frac{1}{\sqrt{2}}(us-su)c   &$for$~\Xi_c^+,\\
        \frac{1}{\sqrt{2}}(ds-sd)c   &$for$~\Xi_c^0.
       \end{cases}
\end{equation}
For the charmed baryons belonging to $\mathbf{6}_F$, the symmetric flavor
wave functions can be written as
\begin{equation}
\phi^c_{\mathbf{6}}=\begin{cases}
                             uuc     &$for$~\Sigma^{++}_c,\\
        \frac{1}{\sqrt{2}}(ud+du)c   &$for$~\Sigma_c^{+},\\
                             ddc     &$for$~\Sigma^{0}_c,\\
        \frac{1}{\sqrt{2}}(us+su)c   &$for$~\Xi_c^{'+},\\
        \frac{1}{\sqrt{2}}(ds+sd)c   &$for$~\Xi_c^{'0},\\
                               ssc   &$for$~\Omega_c^0.
       \end{cases}
\end{equation}

Furthermore, the heavy-quark symmetry as an approximation is commonly
adopted for the study of the singly heavy baryons. In the heavy-quark
symmetry limit, the quark model states may  favor the $j$-$j$ coupling scheme~\cite{Roberts:2007ni}:
\begin{eqnarray}
\left|J^P,j\right\rangle = \left|\left\{\left[\left(\ell_\rho \ell_\lambda\right)_L s_{\rho}\right]_js_Q\right\}_{J^P}\right\rangle,
\end{eqnarray}
where $\ell_\rho$ and $\ell_\lambda$ correspond to the quantum numbers of the orbital
angular momentum $\vell_\rho$ within the light diquark and the orbital
angular momentum $\vell_\lambda$  between the light diquark and
the heavy quark, respectively; $s_{\rho}$ and $s_{Q}$ correspond to
quantum numbers of the spins $\mathbf{s}_{\rho}$ and $\mathbf{s}_Q$ of
the light diquark and heavy quark, respectively; $L$ stands for the quantum number of the
total orbital angular momentum $\mathbf{L}=\vell_\rho+\vell_\lambda$;
$j$ is the quantum number of the total angular momentum $\mathbf{j}=\vell_\rho+\vell_\lambda+\mathbf{s}_{\rho}$
of the light diquark, which is conserved in the heavy quark symmetry limit; and $J$ is the
quantum number of the total angular momentum $\mathbf{J}=\mathbf{j}+\mathbf{s}_Q$ of the heavy
baryon system. The parity of the state is determined by $P=(-1)^{\ell_\rho+\ell_\lambda}$.
In the $j$-$j$ coupling scheme, there are five $\lambda$-mode $P$-wave states
belonging to $\mathbf{6}_F$: $|J^P=\frac{1}{2}^-,0\rangle $, $|J^P=\frac{1}{2}^-,1\rangle $,
$|J^P=\frac{3}{2}^-,1\rangle $, $|J^P=\frac{3}{2}^-,2\rangle $,  and $|J^P=\frac{5}{2}^-,2\rangle $.
Their corresponding quantum numbers are displayed in Table~\ref{JJ}.

The states within the $j$-$j$ coupling scheme are linear combinations of the configurations within the
$L$-$S$ coupling scheme, in which the quark model configurations are
constructed by
\begin{equation}
\left|^{2S+1}L_{J}\right\rangle = \left|\left[\left(\ell_\rho \ell_\lambda\right)_L\left(s_{\rho}s_Q\right)_S\right]_{J^P}\right\rangle,
\end{equation}
where $S$ stands for the quantum number of the total spin angular momentum $\mathbf{S}=\mathbf{s}_{\rho}+\mathbf{s}_Q$.
In the $L$-$S$ coupling scheme, there are also five $\lambda$-mode $P$-wave states:
$|1^{2}P_\lambda \frac{1}{2}^-\rangle$, $|1^{2}P_\lambda \frac{3}{2}^-\rangle$,
$|1^{4}P_\lambda \frac{1}{2}^-\rangle$, $|1^{4}P_\lambda \frac{3}{2}^-\rangle$,
and $|1^{4}P_\lambda \frac{5}{2}^-\rangle$.
The relationship between the $j$-$j$ and $L$-$S$ coupling schemes is given by~\cite{Roberts:2007ni}
\begin{eqnarray}\label{Rela}
\left|\left\{\left[\left(\ell_\rho \ell_\lambda\right)_Ls_{\rho}\right]_js_Q\right\}_{J^P}\right\rangle =(-1)^{L+s_\rho+J+\frac{1}{2}}\sqrt{2j+1} \sum_{S}\sqrt{2S+1}  ~~\nonumber\\
\begin{Bmatrix}L &s_\rho &j\\s_Q &J &S \end{Bmatrix} \left|\left[\left(\ell_\rho \ell_\lambda\right)_L\left(s_{\rho}s_Q\right)_S\right]_{J^P}\right\rangle.~~~~~~~~~
\end{eqnarray}

The heavy quark symmetry may suggest that there are configuration mixing between singly heavy baryon states
with the same $J^P$ numbers in the $L$-$S$ coupling scheme. In the heavy quark limit, the mixing angles
are determined by Eq.(\ref{Rela}).  The two $J^P=1/2^-$ states, $|J^P=\frac{1}{2}^-,0\rangle$ and $|J^P=\frac{1}{2}^-,1\rangle$ in
the $j$-$j$ scheme, are mixed states between $|^{2}P_\lambda \frac{1}{2}^-\rangle$ and $|^{4}P_\lambda \frac{1}{2}^-\rangle$ of the $L$-$S$ coupling scheme with a mixing angle of $\phi\simeq35^\circ$. The two $J^P=3/2^-$ states, $|J^P=\frac{3}{2}^-,1\rangle $ and $|J^P=\frac{3}{2}^-,2\rangle $, are mixed states via  $|^{2}P_\lambda \frac{3}{2}^-\rangle$-$|^{4}P_\lambda \frac{3}{2}^-\rangle$ mixing with a relatively small  angle of $\phi\simeq24^\circ$.

\subsection{Mass analysis}

For $\Lambda_Q$, $\Sigma_Q$, and $\Omega_Q$($Q=c/b$) systems containing two light quarks with an equal mass $m_q$
and one heavy quark with a mass $m_Q$, considering the simplified case of the
harmonic oscillator potentials,  the oscillator frequencies $\omega_\rho$ and
$\omega_\lambda$ for the $\lambda$- and $\rho$-mode excitations satisfy the relation~\cite{Zhong:2007gp,Wang:2017kfr,Yoshida:2015tia,Xiao:2020gjo}
\begin{equation}\label{rho}
\frac{\omega_\lambda}{\omega_\rho}= \sqrt{\frac{1}{3}+\frac{2m_q}{3m_Q}}.
\end{equation}
This relation approximately holds for  $\Xi_Q$ ($Q=b,c$) baryon systems as well,
as the masses of $m_{u/d}$ and $m_s$ may be considered to be approximately equal in the SU(3) limit.
From  Eq.~(\ref{rho}),  when $m_Q\gg m_q$,
the $\lambda$-mode excited energy is smaller than that of the $\rho$ mode, $\omega_\lambda<\omega_\rho$.
In the heavy quark limit,  $m_{Q} \to \infty$, there is no mixing between the $\lambda$- and $\rho$-mode excitations~\cite{Yoshida:2015tia,Capstick:1986bm,Roberts:2007ni}.
This is due to the spin-dependent interaction, which causes the mixing, being suppressed by a factor of $1/m_Q$.
For the singly charm and bottom baryons, the $\lambda$- and $\rho$- modes are well separated~\cite{Yoshida:2015tia,Roberts:2007ni}.
For example,  Yoshida \emph{et al.}  showed the probability of the $\lambda$- and $\rho$- modes of $J^P=1/2^-$
for the lowest $\Sigma_Q$ and $\Lambda_Q$ as a function of the heavy quark mass $m_Q$ in
Fig.~10 of their article~\cite{Yoshida:2015tia}. One sees that the lowest state is almost purely
in the $\lambda$- mode at $m_Q\geq 1.5$ GeV. It should be mentioned that in the SU(3) limit,
i.e., $m_{Q}= m_q$, the excited energies of the $\lambda$- and $\rho$- modes degenerate,
$\omega_\lambda=\omega_\rho$; while the $\lambda$- and $\rho$- modes in the light baryon systems are largely
mixed. The relatively small excitation energy of the $\lambda$- mode indicates
that a $\lambda$-mode excitation should be more easily formed than a $\rho$-mode excitation.
This may explain why most of the newly observed
singly heavy baryons, $\Omega_c(X)$, $\Omega_b(X)$, $\Xi_c(X)$, $\Sigma_b(6097)^{\pm}$, and
$\Xi_b(6227)^0$, may favor the $\lambda$-mode excitations, as predicted in the literature~\cite{Wang:2020gkn,Xiao:2020oif,Wang:2018fjm,
Xiao:2020gjo,Wang:2017kfr,Wang:2017hej,Chen:2018vuc,Chen:2018orb,Chen:2017gnu,Lu:2020ivo,
He:2021xrh,Liang:2020hbo,Yang:2018lzg,Aliev:2018vye,Chen:2017sci,Cui:2019dzj,Yang:2020zjl,Chen:2020mpy,Yang:2020zrh,Agaev:2017lip}.

In the $P$-wave $\Sigma_c$ states, there are two $\rho$-mode
excitations: $J^P=1/2^-$ and $J^P=3/2^-$, and five $\lambda$-mode excitations:
$\Sigma_c|J^P=\frac{1}{2}^-,0\rangle $, $\Sigma_c|J^P=\frac{1}{2}^-,1\rangle $,
$\Sigma_c|J^P=\frac{3}{2}^-,1\rangle $, $\Sigma_c|J^P=\frac{3}{2}^-,2\rangle $, and $\Sigma_c|J^P=\frac{5}{2}^-,2\rangle $.
In this study, we  focus on the $\lambda$-mode $P$-wave $\Sigma_c$ baryon states.
There is a high likelihood that they will be discovered in future experiments, as many
$\lambda$-mode-like states have been observed in the $\Omega_c$, $\Xi_c'$, $\Xi'_b$, $\Omega_b$,
and $\Sigma_b$ families in recent years. Many theoretical approaches, such as the relativized quark model~\cite{Capstick:1986bm}, relativistic quark model~\cite{Ebert:2007nw,Ebert:2011kk,Migura:2006ep}, non-relativistic quark model~\cite{Roberts:2007ni,Chen:2016iyi,Shah:2016mig,Shah:2016nxi,Yoshida:2015tia,Jia:2019bkr}, lattice QCD~\cite{Bahtiyar:2020uuj,Perez-Rubio:2015zqb},
QCD sum rules~\cite{Wang:2010it,Zhang:2008iz,Chen:2015kpa}, and more,
have been adopted in the literature to calculate the mass spectrum.
In Table~\ref{sp1}, we display some masses for the $\lambda$-mode $P$-wave $\Sigma_c$
states predicted within the heavy-quark-light-diquark approximation
in both the relativistic quark model~\cite{Ebert:2007nw,Ebert:2011kk} and non-relativistic quark model~\cite{Chen:2016iyi},
using the hypercentral approximation in the non-relativistic quark model~\cite{Shah:2016mig} and
by strictly solving the three body problem in relativized quark model~\cite{Capstick:1986bm}
and non-relativistic quark model~\cite{Yoshida:2015tia}. The masses predicted using
various approaches using different  approximations are
comparable to each other at $\sim2.71-2.84$ GeV.
There is an obvious gap between the $\rho$- and $\lambda$- mode $P$-wave $\Sigma_c$ states.
The masses for the two $\rho$-mode $P$-wave $\Sigma_c$ states
with $J^P=1/2^-$ and $3/2^-$ are predicted to be $\sim 2.85-2.91$ GeV~\cite{Yoshida:2015tia,Capstick:1986bm},
which are $\sim70$ MeV larger than the highest $\lambda$-mode excitation.

%These $\rho$-mode excitations will be leaved
%for future study, since there are no obvious experimental signals at present.

The $\Sigma_c(2800)$ structure observed in experiments~\cite{Mizuk:2004yu,Aubert:2008ax} is just within
the predicted mass range of $\lambda$-mode $1P$-wave excitations.
However, the predicted mass order and mass splitting between the $P$-wave spin multiplets
is different for various models.
Fortunately, one can determine the mass order and mass splitting for the $1P$-wave $\Sigma_c$ states
from the newly observed $\Xi_c(X)$ and $\Omega_c(X)$ states.
In our previous work~\cite{Wang:2017hej}, the $\Omega_c(3000)$, $\Omega_c(3050)$, $\Omega_c(3065)$,
and $\Omega_c(3090)$ resonances were predicted to be the $\lambda$-mode $1P$-wave
states with $J^P=1/2^-$, $J^P=3/2^-$, $J^P=3/2^-$, and $J^P=5/2^-$, respectively.
The newly observed states, $\Xi_c(2923)$, $\Xi_c(2939)$, and $\Xi_c(2965)$, could be the flavor partners of  $\Omega_c(3050)$, $\Omega_c(3065)$, and $\Omega_c(3090)$, respectively~\cite{Wang:2017hej,Wang:2020gkn,Xiao:2020gjo}.

The equal
spacing rule~\cite{GellMann:1962xb,Okubo:1961jc} perfectly holds for the newly observed $\Xi_c(X)$ and
$\Omega_c(X)$ states, i.e.,
\begin{eqnarray}
&&m[\Omega_c(3050)]-m[\Xi_c(2923)^0] \nonumber \\
&\simeq& m[\Omega_c(3066)]-m[\Xi_c(2939)^0] \nonumber\\
&\simeq& m[\Omega_c(3090)]-m[\Xi_c(2965)^0]\simeq 125 ~\mathrm{MeV}. \nonumber
\end{eqnarray}
The equal spacing rule also perfectly holds
for the $J^P=3/2^+$ charmed ground states:
\begin{eqnarray}
&&m[\Omega_c(2770)^0]-m[\Xi_c(2645)^0] \nonumber \\
&\simeq& m[\Xi_c(2645)^0]-m[\Sigma_c(2520)^0]\simeq 125 ~\mathrm{MeV}. \nonumber
\end{eqnarray}
For the $J^P=1/2^+$ charmed ground states, the equal spacing rule also  holds: $m[\Omega_c]-m[\Xi'_c]$$\simeq m[\Xi'_c]-m[\Sigma_c]\simeq 120$ MeV.
Thus, the equal spacing rule is potentially universal for the charmed baryon states.
Based on this, we predict that for the charmed baryon sector, the flavor partners of the four $\Omega_c(X)$ states,
$\Omega_c(3000)$, $\Omega_c(3050)$, $\Omega_c(3065)$, and $\Omega_c(3090)$, are  likely to be
\begin{eqnarray}
\Omega_c(3000)&:& \Xi_{c}(2873)?, \Sigma_{c}(2746)?, \nonumber \\
\Omega_c(3050)&:& \Xi_{c}(2923),\  \Sigma_{c}(2796)?,\nonumber\\
\Omega_c(3065)&:& \Xi_{c}(2939),\  \Sigma_{c}(2813)?,\nonumber\\
\Omega_c(3090)&:& \Xi_{c}(2965),\  \Sigma_{c}(2840)?,\nonumber
\end{eqnarray}
respectively. The states labeled with $``?"$ are yet to be discovered
by current experiments. Finally,  the equal
spacing rule can be further confirmed using other experimentally observed baryon and meson states.
For example, the equal spacing rule holds:

 (i) for the $J^P=3/2^+$ light ground baryon states,
\begin{eqnarray}
&&m[\Omega(1672)]-m[\Xi(1530)] \nonumber \\
&\simeq& m[\Xi(1530)]-m[\Sigma(1385)] \nonumber\\
&\simeq& m[\Sigma(1385)]-m[\Delta(1232)]\simeq  145~ \mathrm{MeV}; \nonumber
\end{eqnarray}

(ii) for the well-established
light unflavored $n\bar{n}$ ($n=u,d$) and $s\bar{s}$ meson states,
\begin{eqnarray}
&&m[\phi(1020)]-m[\omega(782)] \nonumber \\
&\simeq& m[h_1(1380)]-m[h_1(1170)] \nonumber\\
&\simeq& m[f_2(1270)]-m[f_2'(1520)]\simeq  240 ~\mathrm{MeV}; \nonumber
\end{eqnarray}

(iii) and for the $D$ and $D_s$ meson states,
\begin{eqnarray}
&&m[D_s(1968)]-m[D(1865)^0] \nonumber \\
&\simeq& m[D_s^*(2112)]-m[D^*(2007)^0] \nonumber\\
&\simeq& m[D_{s1}(2536)]-m[D_1(2420)^0] \nonumber\\
&\simeq& m[D_{s2}(2573)]-m[D_2(2460)^0]\simeq  105 ~\mathrm{MeV}. \nonumber
\end{eqnarray}

Less is known about the $P$-wave state
$|J^P=\frac{1}{2}^-,0\rangle$ with the only  hint coming from the recent LHCb experiments
with a broad structure $\Xi_c(2880)$ found in the $\Lambda_c^+K^-$ mass spectrum with
a small significance~\cite{Aaij:2020yyt}. This $\Xi_c(2880)$ may be a candidate for
$|J^P=\frac{1}{2}^-,0\rangle$ in the $\Xi_c'$ family. Thus, according to the equal spacing rule~\cite{GellMann:1962xb,Okubo:1961jc},
the masses of the $\Sigma_c|J^P=\frac{1}{2}^-,0\rangle$ and $\Omega_c|J^P=\frac{1}{2}^-,0\rangle$
state are predicted to be 2755 and 3005 MeV, respectively. The two $P$-wave $J^P=1/2^-$ states,
$|J^P=\frac{1}{2}^-,1\rangle$ and $|J^P=\frac{1}{2}^-,0\rangle$, may be largely overlapping.
It is worth mentioning that for the $\lambda$-mode $P$-wave states there is no place
for $\Omega_c(3119)$, which was observed by LHCb~\cite{Aaij:2017nav}.
The $\Omega_c(3119)$ may be a candidate for the $\lambda$-mode $2S$ states with $J^P=1/2^+$ or $J^P=3/2^+$~\cite{Wang:2017hej}.
Combining this assumption with the equal spacing rule, the masses of the $\lambda$-mode $2S$ state of  $\Sigma_c$ and $\Xi_c$
are predicted to be approximately 2869 and 2994 MeV, respectively.

By combining the precise experimental data of $\Omega_c(X)$ and $\Xi_{c}(X)$ with the equal spacing rule~\cite{GellMann:1962xb,Okubo:1961jc}, we can  determine the
masses for the $\Sigma_c(X)$ states model-independently as corresponding flavor partners of $\Omega_c(X)$ and $\Xi_{c}(X)$.
Then, according to our previous analysis of the strong decays and masses, the newly observed resonances
$\Omega_c(3000)$, $\Omega_c(3050)$/$\Xi_c(2923)$, $\Omega_c(3065)$/$\Xi_c(2939)$, and $\Omega_c(3090)$/$\Xi_c(2965)$
can be naturally explained as the $\lambda$-mode $1P$-wave states with $J^P=1/2^-$, $J^P=3/2^-$, $J^P=3/2^-$, and $J^P=5/2^-$, respectively~\cite{Wang:2017hej,Wang:2020gkn,Xiao:2020gjo,Wang:2017kfr}. Finally, combining the masses
determined  from the equal spacing rule with  possible configuration arrangements based on our previous analysis, we predict a mass spectrum
for the $\lambda$-mode $P$-wave $\Sigma_c$ states, which are summarized in Table~\ref{sp1}. For clarity and comparison, we also plot the mass spectra of the $\lambda$-mode $P$-wave $\Omega_c$, $\Xi_c$ and $\Sigma_c$ baryon states in Fig.~\ref{mass}.

Finally, it should be emphasized that the equal spacing rule perfectly
holds for the $1S$-wave ground charmed baryon states $\Sigma_c^{(*)}$,
$\Xi_c'^{(*)}$ and $\Omega_c^{(*)}$, which is illustrated in Fig.~\ref{mass}.
Considering the newly observed $\Omega_c(X)$ and $\Xi_{c}(X)$ resonances as the
$\lambda$-mode $1P$-wave states, from Fig.~\ref{mass} we observe that
the equal spacing rule also  holds for these higher excitations.
This indicates that the equal spacing rule should hold for all  $1P$-wave
charmed states. Thus, combining the equal spacing rule with the observations,
a fairly reliable and precise prediction of the $P$-wave $\Sigma_c$ states in
a certain mass region can be obtained, as displayed  in Fig.~\ref{mass}. The masses obtained
with the equal spacing rule are comparable with the quark model
predictions~\cite{Ebert:2007nw,Ebert:2011kk,Chen:2016iyi,Capstick:1986bm,
Yoshida:2015tia}, although a precise prediction
cannot be obtained by any quark models due to their typical uncertainties
of approximately 10s to 100 MeV.

\begin{table}[!htb]
\begin{center}
\caption{ \label{sigemac} Strong decay properties of the $\lambda$-mode $P$-wave $\Sigma_c$ states.
The $\Sigma_c$ states are denoted by $|J^P,j\rangle$ in
the $j$-$j$ coupling scheme. The units of the partial widths $\Gamma_{i}$ and masses of the resonances are both MeV. }
%\footnotesize
\begin{tabular}{c|cccccccccccccccccc}
\hline\hline
                  ~~~~State   & $|J^P,j\rangle$       &Channel         &$\Gamma_{i}$     &$\mathcal{B}_i$  \\
\hline
$\Sigma_c(2746)$          &$|J^P=\frac{1}{2}^-,1\rangle$  &$\Lambda_c\pi$         &$\cdot\cdot\cdot$ &$\cdot\cdot\cdot$            \\
                     &                                          &$\Sigma_c\pi$   &$28.06$            &98.00\%      \\
                     &                                          &$\Sigma_c^*\pi$ &$0.47$            & 2.00\%      \\
                     &                                          &total           &$28.53$               &                         \\
       \hline
$\Sigma_c(2755)$          &$|J^P=\frac{1}{2}^-,0\rangle$   &$\Lambda_c\pi$            &$15.23$  &100.00\%          &    \\
                     &                                          &$\Sigma_c\pi$   &$\cdot\cdot\cdot$ &$\cdot\cdot\cdot$     \\
                     &                                          &$\Sigma_c^*\pi$ &$\cdot\cdot\cdot$ &$\cdot\cdot\cdot$    \\
                     &                                          &total           &$15.23$ &                                      \\
        \hline
$\Sigma_c(2796)$          &$|J^P=\frac{3}{2}^-,1\rangle$   &$\Lambda_c\pi$  &$\cdot\cdot\cdot$   &$\cdot\cdot\cdot$        \\
                     &                                         &$\Sigma_c\pi$    &$3.21$   &10.43\%                           \\
                     &                                         &$\Sigma_c^*\pi$  &$27.57$  &89.57\%                              \\
                     &                                         &total            &$30.78$  &                                \\
          \hline
$\Sigma_c(2813)$         &$|J^P=\frac{3}{2}^-, 2\rangle$  &$\Lambda_c\pi$        &$30.59$  &75.42\%            &  \\
                     &                                          &$\Sigma_c\pi$   &$7.53$  &18.57\%                               \\
                     &                                          &$\Sigma_c^*\pi$ &$2.44$  &6.01\%                            \\
                     &                                          &total           &$40.56$  &           \\
         \hline
$\Sigma_c(2840)$          &$|J^P=\frac{5}{2}^-,2\rangle$  &$\Lambda_c\pi$           &$37.63$  &73.96\%                                    \\
        &                                                         &$\Sigma_c\pi$    &$4.90$   &9.63\%                             \\
        &                                                         &$\Sigma_c^*\pi$  &$8.35$   &16.41\%                               \\
        &                                                         &total            &$50.88$  &                                       \\
\hline\hline
\end{tabular}
\end{center}
\end{table}

\section{Strong decay analysis}\label{STD}

\subsection{The model}

Combining with the masses of the
$P$-wave $\Sigma_c$ states, the decay properties   can provide crucial references in searching for them in future experiments.
In this study, we reinvestigate the strong decay properties of $1P$-wave $\Sigma_c$ baryons with the
chiral quark model~\cite{Manohar:1983md}, which has been successfully applied to the strong decays of heavy-light mesons,  and charmed
and strange baryons~\cite{Zhong:2008kd,Zhong:2010vq,Zhong:2009sk,
Zhong:2007gp,Liu:2012sj,Xiao:2013xi,Wang:2017hej,Xiao:2014ura,Xiao:2017udy,Yao:2018jmc,Wang:2020gkn,
Xiao:2020oif,Wang:2018fjm,Xiao:2020gjo,Wang:2019uaj,Wang:2017kfr,Liu:2019wdr,Xiao:2018pwe}.
In this model, the nonrelativistic transition operator for a strong decay process by emitting a
pseudoscalar meson is adopted as~\cite{Zhao:2002id,Li:1994cy,Li:1997gd}
\begin{eqnarray}\label{non-relativistic-expans}
H^{nr}_{m}&=&\sum_j\Big\{\frac{\omega_m}{E_f+M_f}\vsig_j\cdot
\textbf{P}_f+ \frac{\omega_m}{E_i+M_i}\vsig_j \cdot
\textbf{P}_i \nonumber\\
&&-\vsig_j \cdot \textbf{q} +\frac{\omega_m}{2\mu_q}\vsig_j\cdot
\textbf{p}'_j\Big\}I_j e^{-i\mathbf{q}\cdot \mathbf{r}_j},
\end{eqnarray}
where $(E_i,~M_i,~\mathbf{P}_i)$ and $(E_f,~M_f,~\mathbf{P}_f)$ denote the energy, mass, and three-vector momentum of the initial and final baryons, respectively; $\omega_m$ and $\mathbf{q}$ represent the energy and three-vector momentum of the final light pseudoscalar meson; $\mathbf{\sigma}_j$ is the Pauli spin vector on the $j$th quark; $\mathbf{p}'_j (=\mathbf{p}_j-(m_j/M)\mathbf{P}_{\text{c.m.}})$ denotes the internal momentum of the $j$th quark in the baryon rest frame;  and $\mu_q$ is a reduced mass  expressed as $1/\mu_q=1/m_j+1/m'_j$. The isospin operator $I_j$ associated with $\pi$  mesons has been defined in Refs.~\cite{Zhao:2002id,Li:1994cy,Li:1997gd}.
Using the wave functions for the initial and final baryon states,
the transition amplitude $\mathcal{M}$ of a decay process can be determined. For simplicity,
 harmonic oscillator wave functions are adopted for the initial and final baryon states in our calculations.

The partial decay width for a decay process can be calculated using~\cite{Zhong:2007gp,Zhong:2008kd}
\begin{equation}\label{dww}
\Gamma=\left(\frac{\delta}{f_m}\right)^2\frac{(E_f +M_f)|\mathbf{q}|}{4\pi
M_i}\frac{1}{2J_i+1}\sum_{J_{iz}J_{fz}}|\mathcal{M}_{J_{iz},J_{fz}}|^2,
\end{equation}
where $J_{iz}$ and $J_{fz}$ are the third components of the total angular momenta of the initial and final baryons, respectively,
and $\delta$ is a global parameter accounting for the strength of the quark-meson couplings. Here, we fix its value to the same as that in Refs.~\cite{Zhong:2007gp,Zhong:2008kd}, i.e., $\delta=0.557$.

For the calculations, the masses of  well-established hadrons were taken from the Particle Data Group~\cite{Tanabashi:2018oca} and
the standard quark model parameters have been  determined  previously~\cite{Wang:2017kfr}.
For the harmonic oscillator space-wave functions, $\Psi^n_{lm}=R_{nl}Y_{lm}$, the harmonic oscillator parameter $\alpha_{\rho}$
in the wave functions for a $uu/ud/dd$ system is taken as $\alpha_{\rho}=400$ MeV.
Another harmonic oscillator parameter $\alpha_{\lambda}$ is related to $\alpha_{\rho}$
by  $\alpha_{\lambda}=[3m_c/(2m_q+m_c)]^{1/4}\alpha_{\rho}$, where $m_q$
and $m_c$ denote the light $u/d$ quark mass and heavy charmed $c$ quark mass, respectively.
Here, we take $m_q=330$ MeV and $m_c=1480$ MeV.

\subsection{Results and discussions}

Using the predicted masses of the $\lambda$-mode $P$-wave $\Sigma_c$ states from
the mass spectrum analysis in the above section, their strong decay
properties are calculated using the chiral quark model.
The predicted results are display in Table~\ref{sigemac}.

The two $j=1$ states, $\Sigma_c(2746)|J^P=\frac{1}{2}^-,1\rangle$ and
$\Sigma_c(2796)|J^P=\frac{3}{2}^-,1\rangle$, should have very small  decay rates into the
$\Lambda_c \pi$ final state, as this decay mode  is forbidden in
the heavy quark symmetry limit. The $\Sigma_c(2746)|J^P=\frac{1}{2}^-,1\rangle$ and
$\Sigma_c(2796)|J^P=\frac{3}{2}^-,1\rangle$ states have a comparable width of
$\Gamma\simeq 30$ MeV, and predominantly decay into $\Sigma_c\pi$
and $\Sigma_c^*\pi$, respectively. To established these two
states, the $\Lambda_c\pi\pi$ ($\Sigma_c^{(*)}\pi\to \Lambda_c\pi\pi$) invariant
mass spectrum should be observed in future experiments.

The two $j=2$ states, $\Sigma_c(2813)|J^P=\frac{3}{2}^-,2\rangle$ and
$\Sigma_c(2840)|J^P=\frac{5}{2}^-,2\rangle$, predominantly decay into
the $\Lambda_c\pi$ channel with a comparable branching fraction
of $\sim 75\%$. Additionally,  they have relatively large decay rates into $\Sigma_c\pi$
and $\Sigma_c^*\pi$ final states, with  branching fractions of
$\sim 10\%$. The decay widths for the $\Sigma_c(2813)|J^P=\frac{3}{2}^-,2\rangle$ and
$\Sigma_c(2840)|J^P=\frac{5}{2}^-,2\rangle$ states are predicted to be
$\Gamma\simeq 41$ and $51$ MeV, respectively.
These two states might contribute to the
 $\Sigma_c(2800)$ structure observed in the $\Lambda_c\pi$ invariant
mass spectrum. It may be difficult to distinguish
the $\Sigma_c(2813)|J^P=\frac{3}{2}^-,2\rangle$ and
$\Sigma_c(2840)|J^P=\frac{5}{2}^-,2\rangle$ states from the $\Lambda_c\pi$ invariant
mass spectrum due to their highly overlapping masses, which will
be discussed later. Fortunately, as  these two highly overlapping states
have different $J^P$ numbers, they may be separated by measuring
the angular distributions.

The $j=0$ state, $\Sigma_c(2755)|J^P=\frac{1}{2}^-,0\rangle$, may be very
narrow state with a width of $\Gamma\simeq 15$ MeV. Its decays
are likely saturated by the $\Lambda_c\pi$ channel, and the $\Sigma_c\pi$
and $\Sigma_c^*\pi$ decay modes are forbidden in
the heavy quark symmetry limit. It might be interesting
to search for this narrow state in the $\Lambda_c\pi$ channel.

\begin{figure}[htbp]
\centering \epsfxsize=8.0cm \epsfbox{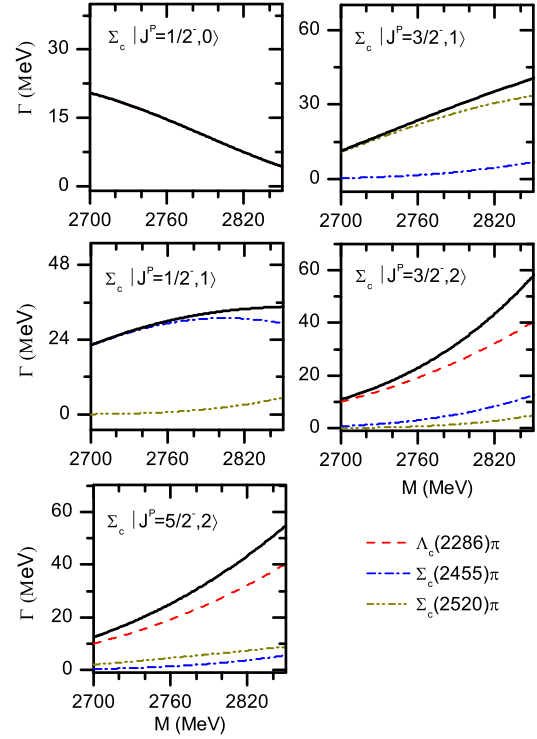}  \vspace{-0.1 cm}
\caption{Strong decay properties of the $\lambda$-mode $1P$-wave $\Sigma_c$ states as functions
of the mass. }\label{sigemaC1}
\end{figure}

Considering the case that the masses of the
$\lambda$-mode $P$-wave $\Sigma_c$ states may be out of our predictions,
in Fig.~\ref{sigemaC1} we plot the strong decay properties as functions
of the mass within a possible region. The sensitivities of the decay
properties to the mass can be clearly seen from the figure.
The uncertainties of the mass for the $\lambda$-mode $P$-wave $\Sigma_c$ states
cannot affect our main conclusion.

Finally,  the strong decay properties
of the two $\rho$-mode $1P$-wave $\Sigma_c$ states with $J^P=\frac{1}{2}^-$ and
$J^P=\frac{3}{2}^-$ are also determined in this study. The $\rho$-mode $1P$-wave states do not
overlap with the $\lambda$-mode $1P$-wave states according to the quark model predictions
~\cite{Yoshida:2015tia,Capstick:1986bm}.
Their masses were predicted to be $M=2909$ and 2910 MeV in a recent study~\cite{Yoshida:2015tia}, which is approximately 70 MeV larger than that of the highest $\lambda$-mode
$P$-wave state, $\Sigma_c(2840)|J^P=\frac{5}{2}^-,2\rangle$.
Additionally, the $\Lambda_c\pi$ decay mode is forbidden for the two $\rho$-mode $1P$-wave $\Sigma_c$ states.
Taking the mass predicted in Ref.~\cite{Yoshida:2015tia}, we find the $J^P=\frac{1}{2}^-$
$\rho$-mode state has a width of $\Gamma\simeq 78$ MeV, and predominantly  decays into
$\Sigma_c\pi$ and $\Sigma_c^*\pi$ with branching fractions of $\sim 65 \%$ and $\sim 35 \%$, respectively,
while the $J^P=\frac{3}{2}^-$ state is  relatively broad, with a width of $\Gamma\simeq 90$ MeV,
and predominantly decays into $\Sigma_c\pi$ and $\Sigma_c^*\pi$ with branching fractions of $\sim 25 \%$ and $\sim 75 \%$, respectively.

\begin{figure}[htbp]
\begin{center}
\centering  \epsfxsize=9.0cm \epsfbox{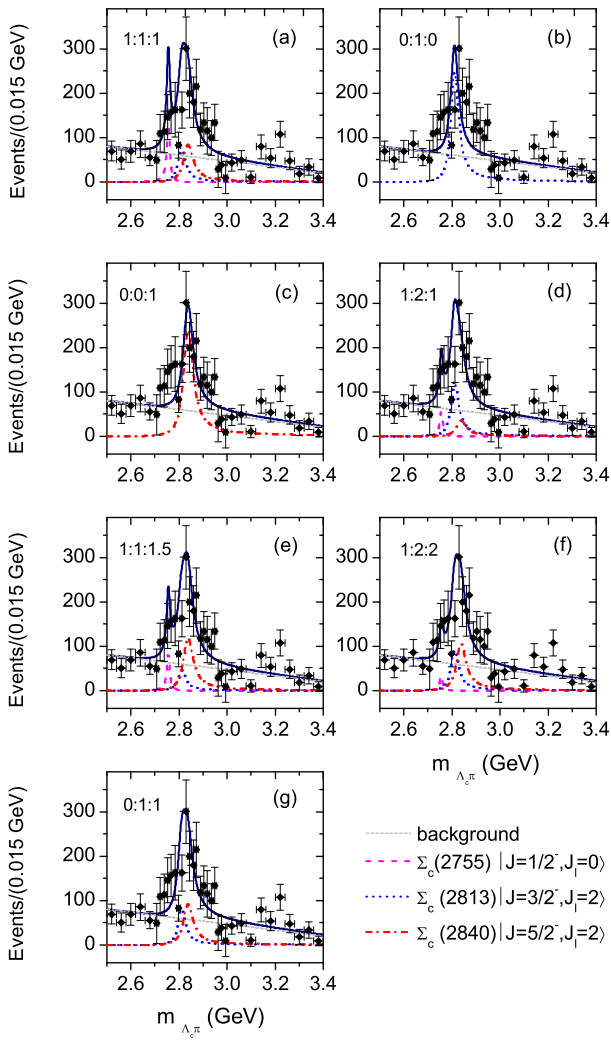}
\vspace{-2.0cm}\caption{The $\Lambda_c \pi$ invariant mass spectrum measured at $BABAR$~\cite{Aubert:2008ax}
(solid squares) compared to the theoretical description with three possible $\lambda$-mode $1P$-wave
states, $\Sigma_c(2746)$, $\Sigma_c(2813)$, and $\Sigma_c(2840)$.
The figures (a)-(e) show seven cases with different resonance production rate ratios, $C_{\Sigma_c(2755)1/2^-}:C_{\Sigma_c(2813)3/2^-}:C_{\Sigma_c(2840)5/2^-}=1:1:1/0:1:0/0:0:1/1:2:1/1:1:1.5/1:2:2/0:1:1$, respectively.} \label{nihe1}
\end{center}
\end{figure}

\section{invariant mass spectrum analysis}\label{IVR}

To determine the contributions of the $P$-wave $\Sigma_c$ states
to  the experimentally observed $\Sigma_c(2800)$ structure, we further analyze the
$\Lambda_c\pi$ invariant mass spectrum measured by $BABAR$~\cite{Aubert:2008ax}.
In our analysis, we adopt a relativistic Breit-Wigner function to describe the event distribution:
\begin{equation}\label{dww}
\frac{dN}{dm}\propto \left|f+\sum_R\frac{ C_R\mathcal{A}_R(m)\sqrt{\Phi (m)}}{m^2-m_R^2+i m_R\Gamma_R(m)}\right|^2,
\end{equation}
where $m$ and $m_{R}$ are the invariant mass and the resonance mass, respectively,
$\mathcal{A}_R(m)$ is the resonance strong decay amplitude for the $\Lambda_c \pi$ channel,
and $\Gamma_R(m)$ denotes the energy-dependent total decay width of a resonance.
The decay width of resonance is considered to be saturated by the
two-body OZI-allowed strong decay modes. Thus, as an approximation,  $\Gamma^R(m)$
is a sum of the energy-dependent partial widths of all two-body OZI-allowed strong decay modes,
 calculated  using the strong decay amplitudes extracted from our chiral quark model.
Additionally, $\Phi (m)$ is strong decay phase space and $f$ represents the background contributions.
In this study, a linear background, $f=\sqrt{a +b m}$, is adopted, where $a\simeq248.0$ MeV
and $b\simeq-67.0$, which were determined by fitting the backgrounds taken in Ref.~\cite{Aubert:2008ax}.
Finally, $\{C_R\}$ is a free parameter set related to the resonance production rates.

According to our strong decay analysis, three $P$-wave $\Sigma_c$
states, $\Sigma_c(2755)|J^P=\frac{1}{2}^-,0\rangle$, $\Sigma_c(2813)|J^P=\frac{3}{2}^-,2\rangle$ and
$\Sigma_c(2840)|J^P=\frac{5}{2}^-,2\rangle$, predominantly decay into the $\Lambda_c\pi$ channel.
Due to the unknown production rates of these resonances, seven cases with different relative ratios, $C_{\Sigma_c(2755)1/2^-}:C_{\Sigma_c(2813)3/2^-}:C_{\Sigma_c(2840)5/2^-}=1:1:1/0:1:0/0:0:1/1:2:1/1:1:1.5//1:2:2/0:1:1$, are considered to reproduce the $\Lambda_c\pi$ invariant mass spectrum measured by $BABAR$ with our predicted
strong decay properties for the three resonances~\cite{Aubert:2008ax}. Our results are displayed in
Figs.~\ref{nihe1} (a)-(g), respectively. Assuming the three resonances with $J^P=1/2^-,3/2^-,5/2^-$ have
comparable production rates as displayed in Figs.~\ref{nihe1} (a), (d), (e), and (f), the measured $\Lambda_c\pi$
invariant mass spectrum can be  described. In these cases,
$\Sigma_c(2755)|J^P=\frac{1}{2}^-,0\rangle$ should contribute a very narrow peak to the invariant mass spectrum,
while the interferences between $\Sigma_c(2813)|J^P=\frac{3}{2}^-,2\rangle$ and
$\Sigma_c(2840)|J^P=\frac{5}{2}^-,2\rangle$ contribute to the main broad peak at approximately 2.8 GeV.
The invariant mass spectrum may also be explained with the two state interferences between $\Sigma_c(2813)|J^P=\frac{3}{2}^-,2\rangle$ and
$\Sigma_c(2840)|J^P=\frac{5}{2}^-,2\rangle$ [see Fig.~\ref{nihe1} (g)].
However, due to the large uncertainties, the present data cannot exclude the possibility that the $\Sigma_c(2800)$ structure is caused by
a single resonance $\Sigma_c(2813)|J^P=\frac{3}{2}^-,2\rangle$ or $\Sigma_c(2840)|J^P=\frac{5}{2}^-,2\rangle$
[see Figs.~\ref{nihe1} (b) and (c)]. The measurements of the angular distributions
are required to separate these two overlapping states.

According to our analysis of the $\Lambda_c\pi$ invariant mass spectrum,
the $\Sigma_c(2800)$ structure may be caused by two largely overlapping resonances,
$\Sigma_c(2813)|J^P=\frac{3}{2}^-,2\rangle$ and $\Sigma_c(2840)|J^P=\frac{5}{2}^-,2\rangle$,
although the explanation with only one single resonance cannot be excluded.
Evidence of the $\Sigma_c(2755)|J^P=\frac{1}{2}^-,0\rangle$ resonance, as a very narrow
peak, may be seen in the $\Lambda_c\pi$ invariant mass spectrum. More accurate measurements of the $\Lambda_c\pi$
invariant mass spectrum along with the partial wave analysis of the measured
angular distributions are crucial for  establishing $\Sigma_c(2755)|J^P=\frac{1}{2}^-,0\rangle$,
$\Sigma_c(2813)|J^P=\frac{3}{2}^-,2\rangle$, and $\Sigma_c(2840)|J^P=\frac{5}{2}^-,2\rangle$.

\section{Summary}\label{SUM}

In this study, by employing the equal spacing rule, the newly observed $\Xi_c(2923)^0$, $\Xi_c(2939)^0$, and $\Xi_c(2965)^0$ states
appear to be flavor partners of $\Omega_c(3050)$, $\Omega_c(3066)$, and $\Omega_c(3090)$, respectively.
As the flavor partners of four $P$-wave candidates, $\Omega_c(3000)$, $\Omega_c(3050)$,
$\Omega_c(3066)$, and $\Omega_c(3090)$, as suggested in the literature,
four $P$-wave $\Sigma_c$ baryon states, $\Sigma_c(2746)$, $\Sigma_c(2796)$,
$\Sigma_c(2813)$, and $\Sigma_c(2840)$, are predicted using the equal spacing rule.
According to our assignments for the $\Omega_c(X)$ states,
 $\Sigma_c(2746)$, $\Sigma_c(2796)$, $\Sigma_c(2813)$, and $\Sigma_c(2840)$,
may correspond to the $\lambda$-mode $P$-wave states $\Sigma_c|J^P=1/2^-,1\rangle$, $\Sigma_c|J^P=3/2^-,1\rangle$,
$\Sigma_c|J^P=3/2^-,2\rangle$, and $\Sigma_c|J^P=5/2^-,2\rangle$, respectively, in
the heavy quark symmetry limit.

Furthermore, their strong decay properties are predicted using the chiral quark model.
It is found that these $1P$-wave $\Sigma_c$ states have relatively narrow widths
within the range of $\sim 15-50$ MeV.  The $\Sigma_c(2813)|J^P=\frac{3}{2}^-,2\rangle$ and
$\Sigma_c(2840)|J^P=\frac{5}{2}^-,2\rangle$ states  have  comparable decay widths of $\Gamma\sim 40$ MeV
and predominantly decay into the $\Lambda_c \pi$ channel.
The $\Sigma_c|J^P=1/2^-,0\rangle$ state may be a very narrow state with
a width of $\Gamma\sim 15$ MeV, and its decays are nearly saturated by
the $\Lambda_c \pi$ channel.
The $\Sigma_c(2755)|J^P=\frac{1}{2}^-,0\rangle$,
$\Sigma_c(2813)|J^P=\frac{3}{2}^-,2\rangle$, and $\Sigma_c(2840)|J^P=\frac{5}{2}^-,2\rangle$ states
may be established in the $\Lambda_c\pi$ invariant mass spectrum with more accurate
measurements and angular distribution analysis in future experiments. The other two $P$-wave states, $\Sigma_c(2746)|J^P=\frac{1}{2}^-,1\rangle$
and $\Sigma_c(2796)|J^P=\frac{3}{2}^-,1\rangle$, are relatively narrow states with comparable widths
of $\Gamma\sim 30$ MeV,  and they  mainly decay into $\Sigma_c\pi$ and $\Sigma^{*}_c\pi$, respectively.
To establish the existence of these two states, the $\Lambda_c\pi\pi$ ($\Sigma_c^{(*)}\pi\to \Lambda_c\pi\pi$) invariant
mass spectrum is worth observing in future experiments.

Then,  using our predicted decay amplitudes, the $\Lambda_c\pi$
invariant mass spectrum measured by $BABAR$ is further analyzed,
which improves our understanding of the nature of $\Sigma_c(2800)$.
It is found that the $\Sigma_c(2800)$ structure can be explained with two largely overlapping resonances,
$\Sigma_c(2813)|J^P=\frac{3}{2}^-,2\rangle$ and $\Sigma_c(2840)|J^P=\frac{5}{2}^-,2\rangle$,
although the explanation with only one single resonance cannot be excluded. If the production rate of $\Sigma_c|J^P=1/2^-,0\rangle$ is
comparable with that of $\Sigma_c(2813)|J^P=3/2^-,2\rangle$ and $\Sigma_c(2840)|J^P=5/2^-,2\rangle$,
a narrow peak will be  observed in the $\Lambda_c \pi$ mass spectrum
at approximately  $2.8$ GeV .

Finally,  the equal spacing rule appears to perfectly
hold for the charmed baryon states. The mass spectrum of the four $P$-wave states,
$\Sigma_c(2746)$, $\Sigma_c(2796)$, $\Sigma_c(2813)$, and $\Sigma_c(2840)$,
is approximately extracted model-independently  by combining the recent observations of the charmed baryon, $\Omega_c(X)$
and $\Xi_c(X)$, with the equal spacing rule. The extracted masses are in the range of
the quark model predictions, and should be more precise than the
quark model predictions due to the highly precise measurements from the LHCb Collaboration.
A reliable determination of the mass spectrum alongside  our detailed analysis
of the decay properties for these $P$-wave $\Sigma_c$ states should
provide useful references for  observations in future experiments.

\section*{Acknowledgements }

This work is supported by the National Natural
Science Foundation of China under Grants No. U1832173,  No. 11775078 and  Scientific and Technological Innovation Programs of Higher Education Institutions of Shanxi Province, China under Grant No.2020L0617.

%\end{spacing}


\begin{thebibliography}{99}

%1\cite{Aaij:2017nav}{Yelton:2017qxg}{Aaij:2018tnn}{Aaij:2018yqz}{Aaij:2019amv}{Aaij:2017ueg}
\bibitem{Aaij:2017nav}
  R.~Aaij {\it et al.} (LHCb Collaboration),
%  Observation of five new narrow $\Omega_c^0$ states decaying to $\Xi_c^+ K^-$,
  Phys.\ Rev.\ Lett.\  {\bf 118}, 182001 (2017)
  %doi:10.1103/PhysRevLett.118.182001
 % [arXiv:1703.04639 [hep-ex]].
  %%CITATION = doi:10.1103/PhysRevLett.118.182001;%%
  %172 citations counted in INSPIRE as of 06 Jan 2020

%2\cite{Aaij:2018yqz}{Aaij:2019amv}{Aaij:2017ueg}
\bibitem{Aaij:2018yqz}
  R.~Aaij {\it et al.} (LHCb Collaboration),
%  Observation of a new $\Xi_b^-$ resonance,
  Phys.\ Rev.\ Lett.\  {\bf 121}, 072002 (2018)
  %doi:10.1103/PhysRevLett.121.072002
  %[arXiv:1805.09418 [hep-ex]].
  %%CITATION = doi:10.1103/PhysRevLett.121.072002;%%
  %43 citations counted in INSPIRE as of 06 Jan 2020

%3\cite{Aaij:2018tnn}{Aaij:2018yqz}{Aaij:2019amv}{Aaij:2017ueg}
\bibitem{Aaij:2018tnn}
  R.~Aaij {\it et al.} (LHCb Collaboration),
%  Observation of Two Resonances in the $\Lambda_b^0 \pi^\pm$ Systems and Precise Measurement of $\Sigma_b^\pm$ and $\Sigma_b^{*\pm}$ properties,
  Phys.\ Rev.\ Lett.\  {\bf 122}, 012001 (2019)
  %doi:10.1103/PhysRevLett.122.012001
  %[arXiv:1809.07752 [hep-ex]].
  %%CITATION = doi:10.1103/PhysRevLett.122.012001;%%
  %29 citations counted in INSPIRE as of 06 Jan 2020

%4\cite{Aaij:2020cex}
\bibitem{Aaij:2020cex}
  R.~Aaij {\it et al.} (LHCb Collaboration),
%  First observation of excited $\Omega_b^-$ states,
  Phys.\ Rev.\ Lett.\  {\bf 124},  082002 (2020)
%  doi:10.1103/PhysRevLett.124.082002
%  [arXiv:2001.00851 [hep-ex]].
  %%CITATION = doi:10.1103/PhysRevLett.124.082002;%%
  %28 citations counted in INSPIRE as of 26 Jan 2021

%5\cite{Aaij:2020yyt}
\bibitem{Aaij:2020yyt}
  R.~Aaij {\it et al.} (LHCb Collaboration),
%  Observation of New $\Xi_c^0$ Baryons Decaying to $\Lambda_c^+ K^-$,
  Phys.\ Rev.\ Lett.\  {\bf 124},  222001 (2020)
%  doi:10.1103/PhysRevLett.124.222001
%  [arXiv:2003.13649 [hep-ex]].
  %%CITATION = doi:10.1103/PhysRevLett.124.222001;%%
  %18 citations counted in INSPIRE as of 26 Jan 2021

%6\cite{Aaij:2020fxj}
\bibitem{Aaij:2020fxj}
  R.~Aaij {\it et al.} (LHCb Collaboration),
%  Observation of a new $\Xi_b^0$ state,
  Phys.\ Rev.\ D {\bf 103}, 012004 (2021)
%  doi:10.1103/PhysRevD.103.012004
%  [arXiv:2010.14485 [hep-ex]].
  %%CITATION = doi:10.1103/PhysRevD.103.012004;%%
  %4 citations counted in INSPIRE as of 19 Feb 2021

%7\cite{Sirunyan:2020gtz}
\bibitem{Sirunyan:2020gtz}
  A.~M.~Sirunyan {\it et al.} (CMS Collaboration),
%  Study of excited $\Lambda_b^0$ states decaying to $\Lambda_b^0\pi^+\pi^-$ in proton-proton collisions at $\sqrt{s}=$ 13 TeV,
  Phys.\ Lett.\ B  803, 135345 (2020)
  %doi:10.1016/j.physletb.2020.135345
  %[arXiv:2001.06533 [hep-ex]].
  %%CITATION = doi:10.1016/j.physletb.2020.135345;%%
  %5 citations counted in INSPIRE as of 14 Apr 2020

%\cite{Aaij:2020rkw}
\bibitem{Aaij:2020rkw}
  R.~Aaij {\it et al.} (LHCb Collaboration),
%  Observation of a new baryon state in the $ {\Lambda}_{\mathrm{b}}^0{\pi}^{+}{\pi}^{-} $ mass spectrum,
  JHEP {\bf 2006}, 136 (2020)
%  doi:10.1007/JHEP06(2020)136
%  [arXiv:2002.05112 [hep-ex]].
  %%CITATION = doi:10.1007/JHEP06(2020)136;%%
  %23 citations counted in INSPIRE as of 17 Jun 2021

%9\cite{Sirunyan:2021vxz}
\bibitem{Sirunyan:2021vxz}
  A.~M.~Sirunyan {\it et al.} (CMS Collaboration),
  Observation of a new excited beauty strange baryon decaying to $\Xi^-_\mathrm{b} \pi^+ \pi^-$,
  arXiv:2102.04524 [hep-ex]
  %%CITATION = ARXIV:2102.04524;%%
  %1 citations counted in INSPIRE as of 19 Feb 2021

%%%%%%%%%%%%%%%%%%%%%%%%%%%%%%%%%%%%%%%%%%%%%%%%%%%%
%45\cite{Roberts:2007ni}
\bibitem{Roberts:2007ni}
  W.~Roberts and M.~Pervin,
%  Heavy baryons in a quark model,
  Int.\ J.\ Mod.\ Phys.\ A {\bf 23}, 2817 (2008)
  %doi:10.1142/S0217751X08041219
  %[arXiv:0711.2492 [nucl-th]].
  %%CITATION = doi:10.1142/S0217751X08041219;%%
  %282 citations counted in INSPIRE as of 07 Jan 2020

%44\cite{Capstick:1986bm}
\bibitem{Capstick:1986bm}
  S.~Capstick and N.~Isgur,
%  Baryons in a Relativized Quark Model with Chromodynamics,
  Phys.\ Rev.\ D {\bf 34}, 2809 (1986)
%  [AIP Conf.\ Proc.\  {\bf 132}, 267 (1985)].
%  doi:10.1103/PhysRevD.34.2809, 10.1063/1.35361
  %%CITATION = doi:10.1103/PhysRevD.34.2809, 10.1063/1.35361;%%
  %1303 citations counted in INSPIRE as of 26 Jan 2021

%50\cite{Yoshida:2015tia}
\bibitem{Yoshida:2015tia}
  T.~Yoshida, E.~Hiyama, A.~Hosaka, M.~Oka and K.~Sadato,
%  Spectrum of heavy baryons in the quark model,
  Phys.\ Rev.\ D {\bf 92}, 114029 (2015)
%  doi:10.1103/PhysRevD.92.114029
%  [arXiv:1510.01067 [hep-ph]].
  %%CITATION = doi:10.1103/PhysRevD.92.114029;%%
  %120 citations counted in INSPIRE as of 10 Mar 2021

%29\cite{Bijker:2020tns}
\bibitem{Bijker:2020tns}
  R.~Bijker, H.~Garc\'{\i}a-Tecocoatzi, A.~Giachino, E.~Ortiz-Pacheco and E.~Santopinto,
  Masses and decay widths of $\Xi_{c/b}$ and $\Xi^\prime_{c/b}$ baryons,
  arXiv:2010.12437 [hep-ph]
  %%CITATION = ARXIV:2010.12437;%%

%64\cite{Liu:2012sj}
\bibitem{Liu:2012sj}
  L.~H.~Liu, L.~Y.~Xiao and X.~H.~Zhong,
%  Charm-strange baryon strong decays in a chiral quark model,
  Phys.\ Rev.\ D  86, 034024 (2012)
 % doi:10.1103/PhysRevD.86.034024
 % [arXiv:1205.2943 [hep-ph]].
  %%CITATION = doi:10.1103/PhysRevD.86.034024;%%
  %33 citations counted in INSPIRE as of 14 Jan 2020

%\cite{Chen:2021eyk}
\bibitem{Chen:2021eyk}
  B.~Chen, S.~Q.~Luo and X.~Liu,
%  Universal behavior of mass gaps existing in the single heavy baryon family,
  Eur.\ Phys.\ J.\ C {\bf 81}, 474 (2021)
%  doi:10.1140/epjc/s10052-021-09234-1
%  [arXiv:2101.10806 [hep-ph]].
  %%CITATION = doi:10.1140/epjc/s10052-021-09234-1;%%

%25\cite{Xiao:2020gjo}
\bibitem{Xiao:2020gjo}
  L.~Y.~Xiao and X.~H.~Zhong,
%  Toward establishing the low-lying $P$-wave $\Sigma_b$ states,
  Phys.\ Rev.\ D {\bf 102}, 014009 (2020)
%  doi:10.1103/PhysRevD.102.014009
%  [arXiv:2004.11106 [hep-ph]].
  %%CITATION = doi:10.1103/PhysRevD.102.014009;%%
  %2 citations counted in INSPIRE as of 02 Feb 2021

%26\cite{Wang:2017kfr}
\bibitem{Wang:2017kfr}
  K.~L.~Wang, Y.~X.~Yao, X.~H.~Zhong and Q.~Zhao,
%  Strong and radiative decays of the low-lying $S$- and $P$-wave singly heavy baryons,
  Phys.\ Rev.\ D  96, 116016 (2017)
  %doi:10.1103/PhysRevD.96.116016
  %[arXiv:1709.04268 [hep-ph]].
  %%CITATION = doi:10.1103/PhysRevD.96.116016;%%
  %34 citations counted in INSPIRE as of 07 Jan 2020

%13\cite{Wang:2017vnc}
\bibitem{Wang:2017vnc}
  W.~Wang and R.~L.~Zhu,
%  Interpretation of the newly observed $\Omega_c^0$ resonances,
  Phys.\ Rev.\ D {\bf 96}, 014024 (2017)
%  doi:10.1103/PhysRevD.96.014024
%  [arXiv:1704.00179 [hep-ph]].
  %%CITATION = doi:10.1103/PhysRevD.96.014024;%%
  %61 citations counted in INSPIRE as of 28 Dec 2020

%12\cite{Chen:2017gnu}
\bibitem{Chen:2017gnu}
  B.~Chen and X.~Liu,
%  New $\Omega_c^0$ baryons discovered by LHCb as the members of $1P$ and $2S$ states,
  Phys.\ Rev.\ D {\bf 96},  094015 (2017)
%  doi:10.1103/PhysRevD.96.094015
%  [arXiv:1704.02583 [hep-ph]].
  %%CITATION = doi:10.1103/PhysRevD.96.094015;%%
  %48 citations counted in INSPIRE as of 28 Dec 2020

%23\cite{Padmanath:2017lng}
\bibitem{Padmanath:2017lng}
  M.~Padmanath and N.~Mathur,
%  Quantum Numbers of Recently Discovered $\Omega^{0}_{c}$ Baryons from Lattice QCD,
  Phys. Rev. Lett. 119, 042001 (2017)
  %doi:10.1103/PhysRevLett.119.042001
  %[arXiv:1704.00259 [hep-ph]].
  %%CITATION = doi:10.1103/PhysRevLett.119.042001;%%
  %51 citations counted in INSPIRE as of 05 Apr 2020

%14\cite{Karliner:2017kfm}  rho mode
\bibitem{Karliner:2017kfm}
  M.~Karliner and J.~L.~Rosner,
%  Very narrow excited $\Omega_c$ baryons,
  Phys.\ Rev.\ D {\bf 95}, 114012 (2017)
%  doi:10.1103/PhysRevD.95.114012
%  [arXiv:1703.07774 [hep-ph]].
  %%CITATION = doi:10.1103/PhysRevD.95.114012;%%
  %75 citations counted in INSPIRE as of 28 Dec 2020

%57\cite{Wang:2017zjw} rho mode
\bibitem{Wang:2017zjw}
  Z.~G.~Wang,
%  Analysis of $\Omega _c(3000)$ , $\Omega _c(3050)$ , $\Omega _c(3066)$ , $\Omega _c(3090)$ and $\Omega _c(3119)$ with QCD sum rules,
  Eur.\ Phys.\ J.\ C {\bf 77}, 325 (2017)
%  doi:10.1140/epjc/s10052-017-4895-5
%  [arXiv:1704.01854 [hep-ph]].
  %%CITATION = doi:10.1140/epjc/s10052-017-4895-5;%%
  %47 citations counted in INSPIRE as of 08 Jun 2021

%58\cite{Aliev:2017led}  rho mode
\bibitem{Aliev:2017led}
  T.~M.~Aliev, S.~Bilmis and M.~Savci,
%  Are the new excited $\Omega_c$ baryons negative parity states?,
  Mod.\ Phys.\ Lett.\ A {\bf 35}, 1950344 (2020)
%  doi:10.1142/S0217732319503449
%  [arXiv:1704.03439 [hep-ph]].
  %%CITATION = doi:10.1142/S0217732319503449;%%
  %28 citations counted in INSPIRE as of 08 Jun 2021

%%%%%%%%%%%%%%%%%%%%%%%% Mass %%%%%%%%%%%%%%%%%%%

%49\cite{Shah:2016nxi}
\bibitem{Shah:2016nxi}
  Z.~Shah, K.~Thakkar, A.~K.~Rai and P.~C.~Vinodkumar,
%  Mass spectra and Regge trajectories of $\Lambda_{c}^{+}$, $\Sigma_{c}^{0}$, $\Xi_{c}^{0}$ and $\Omega_{c}^{0}$ baryons,
  Chin.\ Phys.\ C {\bf 40}, 123102 (2016)
 % doi:10.1088/1674-1137/40/12/123102
 % [arXiv:1609.08464 [nucl-th]].
  %%CITATION = doi:10.1088/1674-1137/40/12/123102;%%
  %62 citations counted in INSPIRE as of 10 Mar 2021

%39\cite{Chen:2015kpa}
\bibitem{Chen:2015kpa}
  H.~X.~Chen, W.~Chen, Q.~Mao, A.~Hosaka, X.~Liu and S.~L.~Zhu,
%  P-wave charmed baryons from QCD sum rules,
  Phys.\ Rev.\ D {\bf 91}, 054034 (2015)
%  doi:10.1103/PhysRevD.91.054034
%  [arXiv:1502.01103 [hep-ph]].
  %%CITATION = doi:10.1103/PhysRevD.91.054034;%%
  %79 citations counted in INSPIRE as of 27 Jan 2021

%46\cite{Bahtiyar:2020uuj}
\bibitem{Bahtiyar:2020uuj}
  H.~Bahtiyar, K.~U.~Can, G.~Erkol, P.~Gubler, M.~Oka and T.~T.~Takahashi,
%  Charmed baryon spectrum from lattice QCD near the physical point,
  Phys.\ Rev.\ D {\bf 102}, 054513 (2020)
%  doi:10.1103/PhysRevD.102.054513
%  [arXiv:2004.08999 [hep-lat]].
  %%CITATION = doi:10.1103/PhysRevD.102.054513;%%
  %8 citations counted in INSPIRE as of 10 Mar 2021

%\cite{Perez-Rubio:2015zqb}
\bibitem{Perez-Rubio:2015zqb}
P.~P\'erez-Rubio, S.~Collins and G.~S.~Bali,
%Charmed baryon spectroscopy and light flavor symmetry from lattice QCD,
Phys. Rev. D \textbf{92}, 034504 (2015)
%doi:10.1103/PhysRevD.92.034504
%[arXiv:1503.08440 [hep-lat]].
%90 citations counted in INSPIRE as of 12 Aug 2021

%36\cite{Ebert:2007nw}
\bibitem{Ebert:2007nw}
  D.~Ebert, R.~N.~Faustov and V.~O.~Galkin,
%  Masses of excited heavy baryons in the relativistic quark model,
  Phys.\ Lett.\ B {\bf 659}, 612 (2008)
 % doi:10.1016/j.physletb.2007.11.037
  %[arXiv:0705.2957 [hep-ph]].
  %%CITATION = doi:10.1016/j.physletb.2007.11.037;%%
  %166 citations counted in INSPIRE as of 07 Jan 2020

%42\cite{Ebert:2011kk}
\bibitem{Ebert:2011kk}
  D.~Ebert, R.~N.~Faustov and V.~O.~Galkin,
%  Spectroscopy and Regge trajectories of heavy baryons in the relativistic quark-diquark picture,
  Phys.\ Rev.\ D {\bf 84}, 014025 (2011)
  %doi:10.1103/PhysRevD.84.014025
  %[arXiv:1105.0583 [hep-ph]].
  %%CITATION = doi:10.1103/PhysRevD.84.014025;%%
  %134 citations counted in INSPIRE as of 07 Jan 2020

%27\cite{Yamaguchi:2014era}
\bibitem{Yamaguchi:2014era}
  Y.~Yamaguchi, S.~Ohkoda, A.~Hosaka, T.~Hyodo and S.~Yasui,
% Heavy quark symmetry in multihadron systems,
  Phys. Rev. D  91, 034034 (2015)
 % doi:10.1103/PhysRevD.91.034034
 % [arXiv:1402.5222 [hep-ph]].
  %%CITATION = doi:10.1103/PhysRevD.91.034034;%%
  %34 citations counted in INSPIRE as of 03 Apr 2020

%38\cite{Garcilazo:2007eh}
\bibitem{Garcilazo:2007eh}
  H.~Garcilazo, J.~Vijande and A.~Valcarce,
%  Faddeev study of heavy baryon spectroscopy,
  J.\ Phys.\ G {\bf 34}, 961 (2007)
  %doi:10.1088/0954-3899/34/5/014
  %[hep-ph/0703257].
  %%CITATION = doi:10.1088/0954-3899/34/5/014;%%
  %108 citations counted in INSPIRE as of 07 Jan 2020

%51\cite{Wang:2010it}
\bibitem{Wang:2010it}
  Z.~G.~Wang,
%  Analysis of the ${1/2^-}$ and ${3/2^-}$ heavy and doubly heavy baryon states with QCD sum rules,
  Eur.\ Phys.\ J.\ A {\bf 47}, 81 (2011)
%  doi:10.1140/epja/i2011-11081-8
%  [arXiv:1003.2838 [hep-ph]].
  %%CITATION = doi:10.1140/epja/i2011-11081-8;%%
  %61 citations counted in INSPIRE as of 10 Mar 2021

%\cite{Xu:2020ofp}
\bibitem{Xu:2020ofp}
  Y.~J.~Xu, Y.~L.~Liu, C.~Y.~Cui and M.~Q.~Huang,
  P-wave $\Omega_{b}$ states: masses and pole residues,
  arXiv:2010.10697 [hep-ph]
  %%CITATION = ARXIV:2010.10697;%%
  %1 citations counted in INSPIRE as of 12 Jun 2021

 %80\cite{Mao:2015gya}
\bibitem{Mao:2015gya}
  Q.~Mao, H.~X.~Chen, W.~Chen, A.~Hosaka, X.~Liu and S.~L.~Zhu,
%  QCD sum rule calculation for P-wave bottom baryons,
  Phys.\ Rev.\ D {\bf 92}, 114007 (2015)
%  doi:10.1103/PhysRevD.92.114007
%  [arXiv:1510.05267 [hep-ph]].
  %%CITATION = doi:10.1103/PhysRevD.92.114007;%%
  %55 citations counted in INSPIRE as of 10 Jun 2021

%63\cite{Karliner:2020fqe} rho mode
\bibitem{Karliner:2020fqe}
  M.~Karliner and J.~L.~Rosner,
%  Interpretation of excited $\Omega_b$ signals,
  Phys.\ Rev.\ D {\bf 102}, 014027 (2020)
%  doi:10.1103/PhysRevD.102.014027
%  [arXiv:2005.12424 [hep-ph]].
  %%CITATION = doi:10.1103/PhysRevD.102.014027;%%
  %7 citations counted in INSPIRE as of 08 Jun 2021

%61\cite{Wang:2020pri} rho mode
\bibitem{Wang:2020pri}
  Z.~G.~Wang,
%  Analysis of the $\Omega_b(6316)$, $\Omega_b(6330)$, $\Omega_b(6340)$ and $\Omega_b(6350)$ with QCD sum rules,
  Int.\ J.\ Mod.\ Phys.\ A {\bf 35}, 2050043 (2020)
%  doi:10.1142/S0217751X20500438
%  [arXiv:2001.02961 [hep-ph]].
  %%CITATION = doi:10.1142/S0217751X20500438;%%
  %10 citations counted in INSPIRE as of 08 Jun 2021

%\cite{Mutuk:2020rzm}
\bibitem{Mutuk:2020rzm}
  H.~Mutuk,
%  A study of excited $\Omega _b^-$ states in hypercentral constituent quark model via artificial neural network,
  Eur.\ Phys.\ J.\ A {\bf 56}, 146 (2020)
%  doi:10.1140/epja/s10050-020-00161-5
%  [arXiv:2002.03695 [hep-ph]].
  %%CITATION = doi:10.1140/epja/s10050-020-00161-5;%%
  %3 citations counted in INSPIRE as of 12 Jun 2021

%%%%%%%%%%%%%%%%%%%%%%%%%%%%%%%%%%%QCD SR%%%%%%%%%%%%%%%%%%%%%%%%%%%%%%%%%%%%%%%%%%%%%%%%%%%%%

%30\cite{Yang:2020zjl}
\bibitem{Yang:2020zjl}
  H.~M.~Yang, H.~X.~Chen and Q.~Mao,
%  Identifying the $\Xi_c^0$ baryons observed by LHCb as $P$-wave $\Xi_c^\prime$ baryons,
  Phys.\ Rev.\ D {\bf 102}, 114009 (2020)
%  doi:10.1103/PhysRevD.102.114009
%  [arXiv:2004.00531 [hep-ph]].
  %%CITATION = doi:10.1103/PhysRevD.102.114009;%%
  %6 citations counted in INSPIRE as of 26 Jan 2021

%62\cite{Chen:2020mpy}
\bibitem{Chen:2020mpy}
  H.~X.~Chen, E.~L.~Cui, A.~Hosaka, Q.~Mao and H.~M.~Yang,
%  Excited $\Omega _b$ baryons and fine structure of strong interaction,
  Eur.\ Phys.\ J.\ C {\bf 80}, 256 (2020)
%  doi:10.1140/epjc/s10052-020-7824-y
%  [arXiv:2001.02147 [hep-ph]].
  %%CITATION = doi:10.1140/epjc/s10052-020-7824-y;%%
  %12 citations counted in INSPIRE as of 08 Jun 2021

%15\cite{Chen:2017sci}
\bibitem{Chen:2017sci}
  H.~X.~Chen, Q.~Mao, W.~Chen, A.~Hosaka, X.~Liu and S.~L.~Zhu,
%  Decay properties of $P$-wave charmed baryons from light-cone QCD sum rules,
  Phys. Rev. D  95, 094008 (2017)
  %doi:10.1103/PhysRevD.95.094008
  %[arXiv:1703.07703 [hep-ph]].
  %%CITATION = doi:10.1103/PhysRevD.95.094008;%%
  %58 citations counted in INSPIRE as of 05 Apr 2020

%\cite{Yang:2019cvw}
\bibitem{Yang:2019cvw}
  H.~M.~Yang, H.~X.~Chen, E.~L.~Cui, A.~Hosaka and Q.~Mao,
%  Decay properties of $P$-wave bottom baryons within light-cone sum rules,
  Eur. Phys. J. C \textbf{80}, 80 (2020)
%doi:10.1140/epjc/s10052-020-7637-z
%[arXiv:1909.13575 [hep-ph]].
%13 citations counted in INSPIRE as of 10 Aug 2021

%20\cite{Cui:2019dzj}
\bibitem{Cui:2019dzj}
  E.~L.~Cui, H.~M.~Yang, H.~X.~Chen and A.~Hosaka,
%  Identifying the $\Xi_{b}(6227)$ and $\Sigma_{b}(6097)$ as $P$-wave bottom baryons of $J^P = 3/2^-$,
  Phys.\ Rev.\ D {\bf 99},  094021 (2019)
%  doi:10.1103/PhysRevD.99.094021
%  [arXiv:1903.10369 [hep-ph]].
  %%CITATION = doi:10.1103/PhysRevD.99.094021;%%
  %19 citations counted in INSPIRE as of 28 Dec 2020

%\cite{Yang:2020zrh}
\bibitem{Yang:2020zrh}
  H.~M.~Yang and H.~X.~Chen,
%%  $P$-wave bottom baryons of the $SU(3)$ flavor $\mathbf{6}_F$,
  Phys.\ Rev.\ D {\bf 101}, 114013 (2020)
  Erratum: [Phys.\ Rev.\ D {\bf 102}, 079901 (2020)]
%  doi:10.1103/PhysRevD.101.114013, 10.1103/PhysRevD.102.079901
%  [arXiv:2003.07488 [hep-ph]].
  %%CITATION = doi:10.1103/PhysRevD.101.114013, 10.1103/PhysRevD.102.079901;%%
  %9 citations counted in INSPIRE as of 12 Jun 2021


%22\cite{Aliev:2018vye}
\bibitem{Aliev:2018vye}
  T.~M.~Aliev, K.~Azizi, Y.~Sarac and H.~Sundu,
%  Determination of the quantum numbers of $\Sigma_b(6097)^{\pm}$ via their strong decays,
  Phys.\ Rev.\ D {\bf 99},  094003 (2019)
%  doi:10.1103/PhysRevD.99.094003
%  [arXiv:1811.05686 [hep-ph]].
  %%CITATION = doi:10.1103/PhysRevD.99.094003;%%
  %14 citations counted in INSPIRE as of 28 Dec 2020

%10\cite{Agaev:2017lip}
\bibitem{Agaev:2017lip}
  S.~S.~Agaev, K.~Azizi and H.~Sundu,
%  Interpretation of the new $\Omega_c^{0}$ states via their mass and width,
  Eur.\ Phys.\ J.\ C {\bf 77},  395 (2017)
%  doi:10.1140/epjc/s10052-017-4953-z
%  [arXiv:1704.04928 [hep-ph]].
  %%CITATION = doi:10.1140/epjc/s10052-017-4953-z;%%
  %40 citations counted in INSPIRE as of 28 Dec 2020


%%%%%%%%%%%%%%%%%%%%%%%%%%%3P0 model%%%%%%%%%%%%%%%%%%%%%%%%%%%%%%%%%%%%%%%%%%%%%%%%%%%%%%

%60\cite{Liang:2020hbo}
\bibitem{Liang:2020hbo}
  W.~Liang and Q.~F.~L\"{u},
%  Strong decays of the newly observed narrow $\Omega _b$ structures,
  Eur.\ Phys.\ J.\ C {\bf 80} (2020)  198
%  doi:10.1140/epjc/s10052-020-7759-3
%  [arXiv:2001.02221 [hep-ph]].
  %%CITATION = doi:10.1140/epjc/s10052-020-7759-3;%%
  %12 citations counted in INSPIRE as of 08 Jun 2021

%\cite{Santopinto:2018ljf}
\bibitem{Santopinto:2018ljf}
E.~Santopinto, A.~Giachino, J.~Ferretti, H.~Garc\'\i{}a-Tecocoatzi, M.~A.~Bedolla, R.~Bijker and E.~Ortiz-Pacheco,
%The $\Omega_{c}$-puzzle solved by means of quark model predictions,
Eur. Phys. J. C \textbf{79}, 1012 (2019)
%doi:10.1140/epjc/s10052-019-7527-4
%[arXiv:1811.01799 [hep-ph]].
%18 citations counted in INSPIRE as of 10 Aug 2021

%18\cite{Chen:2018vuc}
\bibitem{Chen:2018vuc}
  B.~Chen and X.~Liu,
%  Assigning the newly reported $\Sigma_b(6097)$ as a $P$-wave excited state and predicting its partners,
  Phys.\ Rev.\ D {\bf 98}, 074032 (2018)
%  doi:10.1103/PhysRevD.98.074032
%  [arXiv:1810.00389 [hep-ph]].
  %%CITATION = doi:10.1103/PhysRevD.98.074032;%%
  %23 citations counted in INSPIRE as of 28 Dec 2020

%19\cite{Chen:2018orb}
\bibitem{Chen:2018orb}
  B.~Chen, K.~W.~Wei, X.~Liu and A.~Zhang,
%  Role of newly discovered $\Xi_b(6227)^-$ for constructing excited bottom baryon family,
  Phys.\ Rev.\ D {\bf 98},  031502 (2018)
%  doi:10.1103/PhysRevD.98.031502
%  [arXiv:1805.10826 [hep-ph]].
  %%CITATION = doi:10.1103/PhysRevD.98.031502;%%
  %23 citations counted in INSPIRE as of 28 Dec 2020

%17\cite{Yang:2018lzg}
\bibitem{Yang:2018lzg}
  P.~Yang, J.~J.~Guo and A.~Zhang,
%  Identification of the newly observed $\Sigma_b(6097)^\pm$ baryons from their strong decays,
  Phys.\ Rev.\ D {\bf 99},  034018 (2019)
%  doi:10.1103/PhysRevD.99.034018
%  [arXiv:1810.06947 [hep-ph]].
  %%CITATION = doi:10.1103/PhysRevD.99.034018;%%
  %16 citations counted in INSPIRE as of 28 Dec 2020

%\cite{He:2021xrh}
\bibitem{He:2021xrh}
H.~Z.~He, W.~Liang, Q.~F.~L\"u and Y.~B.~Dong,
%Strong decays of the low-lying bottom strange baryons,
Sci. China Phys. Mech. Astron. \textbf{64},  261012 (2021)
%doi:10.1007/s11433-021-1704-x
%[arXiv:2102.07391 [hep-ph]].
%0 citations counted in INSPIRE as of 10 Aug 2021

 %28\cite{Lu:2020ivo}
\bibitem{Lu:2020ivo}
  Q.~F.~L\"{u},
%  Canonical interpretations of the newly observed $\Xi _c(2923)^0$, $\Xi _c(2939)^0$, and $\Xi _c(2965)^0$ resonances,
  Eur.\ Phys.\ J.\ C {\bf 80},  921 (2020)
%  doi:10.1140/epjc/s10052-020-08488-5
%  [arXiv:2004.02374 [hep-ph]].
  %%CITATION = doi:10.1140/epjc/s10052-020-08488-5;%%
  %8 citations counted in INSPIRE as of 26 Jan 2021


%%%%%%%%%%%%%%%%%%%%%%%%%%%%Chiral quark model %%%%%%%%%%%%%%%%%%%%%%%%%%%%%%%%%%%%%%%%%%%%%%%%%%%%%%%
%21\cite{Wang:2018fjm}
\bibitem{Wang:2018fjm}
  K.~L.~Wang, Q.~F.~L\"{u} and X.~H.~Zhong,
%  Interpretation of the newly observed $\Sigma_b(6097)^{\pm}$ and $\Xi_b(6227)^-$ states as the $P$-wave bottom baryons,
  Phys.\ Rev.\ D {\bf 99}, 014011 (2019)
  %%doi:10.1103/PhysRevD.99.014011
  %[arXiv:1810.02205 [hep-ph]].
  %%CITATION = doi:10.1103/PhysRevD.99.014011;%%
  %11 citations counted in INSPIRE as of 07 Jan 2020

%24\cite{Wang:2020gkn}
\bibitem{Wang:2020gkn}
  K.~L.~Wang, L.~Y.~Xiao and X.~H.~Zhong,
%  Understanding the newly observed $\Xi_c^0$ states through their decays,
  Phys.\ Rev.\ D {\bf 102},  034029 (2020)
%  doi:10.1103/PhysRevD.102.034029
%  [arXiv:2004.03221 [hep-ph]].
  %%CITATION = doi:10.1103/PhysRevD.102.034029;%%
  %9 citations counted in INSPIRE as of 26 Jan 2021

%27\cite{Xiao:2020oif}
\bibitem{Xiao:2020oif}
  L.~Y.~Xiao, K.~L.~Wang, M.~S.~Liu and X.~H.~Zhong,
%  Possible interpretation of the newly observed $\Omega_b$ states,
  Eur.\ Phys.\ J.\ C  80, 279 (2020)
 % doi:10.1140/epjc/s10052-020-7823-z
  %[arXiv:2001.05110 [hep-ph]].
  %%CITATION = doi:10.1140/epjc/s10052-020-7823-z;%%
  %4 citations counted in INSPIRE as of 17 Apr 2020

%16\cite{Wang:2017hej}
\bibitem{Wang:2017hej}
  K.~L.~Wang, L.~Y.~Xiao, X.~H.~Zhong and Q.~Zhao,
%  Understanding the newly observed $\Omega_c$ states through their decays,
  Phys.\ Rev.\ D {\bf 95}, 116010 (2017)
  %doi:10.1103/PhysRevD.95.116010
  %[arXiv:1703.09130 [hep-ph]].
  %%CITATION = doi:10.1103/PhysRevD.95.116010;%%
  %48 citations counted in INSPIRE as of 08 Jan 2020

%11\cite{Cheng:2017ove}
\bibitem{Cheng:2017ove}
  H.~Y.~Cheng and C.~W.~Chiang,
%  Quantum numbers of $\Omega_c$ states and other charmed baryons,
  Phys.\ Rev.\ D {\bf 95}, 094018 (2017)
%  doi:10.1103/PhysRevD.95.094018
%  [arXiv:1704.00396 [hep-ph]].
  %%CITATION = doi:10.1103/PhysRevD.95.094018;%%
  %51 citations counted in INSPIRE as of 28 Dec 2020
%%%%%%%%%%%%%%%%%%%%%%%%%%%%%%%%%%%%%%%%%%%%%%%%%%%%%%%%%%%%%%%%%%%%%%%%%%%%%%%%%%%%%%%%%%%

%\cite{Agaev:2017jyt}
\bibitem{Agaev:2017jyt}
S.~S.~Agaev, K.~Azizi and H.~Sundu,
%On the nature of the newly discovered $\Omega_c^0$ states,
EPL \textbf{118}, 61001 (2017)
%doi:10.1209/0295-5075/118/61001
%[arXiv:1703.07091 [hep-ph]].
%59 citations counted in INSPIRE as of 10 Aug 2021


%\cite{LHCb:2021ptx}
\bibitem{LHCb:2021ptx}
R.~Aaij \textit{et al.} (LHCb Collaboration),
%Observation of excited $\Omega_c^0$ baryons in $\Omega_b^- \to \Xi_c^+ K^-\pi^-$decays,
Phys. Rev. D \textbf{104}, L091102 (2021)
%doi:10.1103/PhysRevD.104.L091102
%[arXiv:2107.03419 [hep-ex]].
%5 citations counted in INSPIRE as of 30 Nov 2021

%\cite{Cheng:2021qpd}
\bibitem{Cheng:2021qpd}
H.~Y.~Cheng,
Charmed Baryon Physics Circa 2021,
[arXiv:2109.01216 [hep-ph]]
%3 citations counted in INSPIRE as of 29 Nov 2021


%\cite{Agaev:2020fut}
\bibitem{Agaev:2020fut}
S.~S.~Agaev, K.~Azizi and H.~Sundu,
%Newly discovered $\Xi _c^{0}$ resonances and their parameters,
Eur. Phys. J. A \textbf{57}, 201 (2021)
%doi:10.1140/epja/s10050-021-00523-7
%[arXiv:2007.00583 [hep-ph]].
%3 citations counted in INSPIRE as of 10 Aug 2021

%%%%%%%%%%%%%%%%%%%%%%%%%%%%%%%%%%%%%%%%%%%%%%%%%%%%%%%%%%%%%%%%%%%%%%%%%%%%%%%%%

%48\cite{Jia:2019bkr}
\bibitem{Jia:2019bkr}
  D.~Jia, W.~N.~Liu and A.~Hosaka,
%  Regge behaviors in orbitally excited spectroscopy of charmed and bottom baryons,
  Phys.\ Rev.\ D {\bf 101}, 034016 (2020)
%  [arXiv:1907.04958 [hep-ph]].
  %%CITATION = ARXIV:1907.04958;%%
  %13 citations counted in INSPIRE as of 10 Mar 2021

%%%%%%%%%%%%%%%%%%%%%%%%%%%%%%%%exotic%%%%%%%%%%%%%%%%%%%%%%%%%%%%%%

%\cite{Wang:2018alb}
\bibitem{Wang:2018alb}
Z.~G.~Wang and J.~X.~Zhang,
%Possible pentaquark candidates: new excited $\Omega _c$ states,
Eur. Phys. J. C \textbf{78}, 503 (2018)
%doi:10.1140/epjc/s10052-018-5989-4
%[arXiv:1804.06195 [hep-ph]].
%13 citations counted in INSPIRE as of 10 Aug 2021

%\cite{Huang:2017dwn}
\bibitem{Huang:2017dwn}
H.~Huang, J.~Ping and F.~Wang,
%Investigating the excited $\Omega^{0}_{c}$ states through $\Xi_{c}K$ and $\Xi^{'}_{c}K$ decay channels,
Phys. Rev. D \textbf{97}, 034027 (2018)
%doi:10.1103/PhysRevD.97.034027
%[arXiv:1704.01421 [hep-ph]].
%43 citations counted in INSPIRE as of 10 Aug 2021

%\cite{Kim:2017jpx}
\bibitem{Kim:2017jpx}
H.~C.~Kim, M.~V.~Polyakov and M.~Prasza\l{}owicz,
%Possibility of the existence of charmed exotica,
Phys. Rev. D \textbf{96}, 014009 (2017)
%doi:10.1103/PhysRevD.96.014009
%[arXiv:1704.04082 [hep-ph]].
%60 citations counted in INSPIRE as of 10 Aug 2021

%\cite{An:2017lwg}
\bibitem{An:2017lwg}
C.~S.~An and H.~Chen,
%Observed $\Omega_{c}^{0}$ resonances as pentaquark states,
Phys. Rev. D \textbf{96}, 034012 (2017)
%doi:10.1103/PhysRevD.96.034012
%[arXiv:1705.08571 [hep-ph]].
%32 citations counted in INSPIRE as of 10 Aug 2021

%\cite{Kim:2017khv}
\bibitem{Kim:2017khv}
H.~C.~Kim, M.~V.~Polyakov, M.~Praszalowicz and G.~S.~Yang,
%Strong decays of exotic and nonexotic heavy baryons in the chiral quark-soliton model,
Phys. Rev. D \textbf{96}, 094021 (2017)
[erratum: Phys. Rev. D \textbf{97}, 039901 (2018)]
%doi:10.1103/PhysRevD.96.094021
%[arXiv:1709.04927 [hep-ph]].
%30 citations counted in INSPIRE as of 10 Aug 2021

%\cite{Montana:2017kjw}
\bibitem{Montana:2017kjw}
G.~Monta\~na, A.~Feijoo and \`A.~Ramos,
%A meson-baryon molecular interpretation for some $\Omega_{c}$ excited states,
Eur. Phys. J. A \textbf{54}, 64 (2018)
%doi:10.1140/epja/i2018-12498-1
%[arXiv:1709.08737 [hep-ph]].
%45 citations counted in INSPIRE as of 10 Aug 2021

%\cite{Debastiani:2017ewu}
\bibitem{Debastiani:2017ewu}
V.~R.~Debastiani, J.~M.~Dias, W.~H.~Liang and E.~Oset,
%Molecular $\Omega_c$ states generated from coupled meson-baryon channels,
Phys. Rev. D \textbf{97}, 094035 (2018)
%doi:10.1103/PhysRevD.97.094035
%[arXiv:1710.04231 [hep-ph]].
%51 citations counted in INSPIRE as of 10 Aug 2021

%\cite{Huang:2018wgr}
\bibitem{Huang:2018wgr}
Y.~Huang, C.~j.~Xiao, Q.~F.~L\"u, R.~Wang, J.~He and L.~Geng,
%Strong and radiative decays of $D\Xi$ molecular state and newly observed $\Omega_c$ states,
Phys. Rev. D \textbf{97}, 094013 (2018)
%doi:10.1103/PhysRevD.97.094013
%[arXiv:1801.03598 [hep-ph]].
%25 citations counted in INSPIRE as of 10 Aug 2021

%\cite{Debastiani:2018adr}
\bibitem{Debastiani:2018adr}
V.~R.~Debastiani, J.~M.~Dias, W.~H.~Liang and E.~Oset,
%$\boldsymbol{\Omega_b^- \to (\Xi_c^+ \, K^-) \, \pi^-}$ and the $\boldsymbol{\Omega_c}$ states,
Phys. Rev. D \textbf{98}, 094022 (2018)
%doi:10.1103/PhysRevD.98.094022
%[arXiv:1803.03268 [hep-ph]].
%3 citations counted in INSPIRE as of 10 Aug 2021

%\cite{Liu:2018bkx}
\bibitem{Liu:2018bkx}
M.~Z.~Liu, T.~W.~Wu, J.~J.~Xie, M.~Pavon Valderrama and L.~S.~Geng,
%$D \Xi$ and $D^* \Xi$ molecular states from one boson exchange,
Phys. Rev. D \textbf{98}, 014014 (2018)
%doi:10.1103/PhysRevD.98.014014
%[arXiv:1805.08384 [hep-ph]].
%16 citations counted in INSPIRE as of 10 Aug 2021

%\cite{Montana:2018edp}
\bibitem{Montana:2018edp}
G.~Monta\~na, \`A.~Ramos and A.~Feijoo,
%Exotic $\Omega_c^0$ baryons from meson-baryon scattering,
J. Phys. Conf. Ser. \textbf{1137}, 012040 (2019)
%doi:10.1088/1742-6596/1137/1/012040
%[arXiv:1812.03890 [hep-ph]].
%0 citations counted in INSPIRE as of 10 Aug 2021

%\cite{Wang:2021cku}
\bibitem{Wang:2021cku}
H.~J.~Wang, Z.~Y.~Di and Z.~G.~Wang,
%Analysis of the excited $\Omega_c$ states as the pentaquark states with QCD sum rules,
Commun. Theor. Phys. \textbf{73}, 035201 (2021)
%doi:10.1088/1572-9494/abc7b1
%0 citations counted in INSPIRE as of 10 Aug 2021

%%%%%%%%%%%%%%%%%%%Omigab%%%%%%%%%%%%%%%%%%%%%%%%%%%%%%%%%%%%%%%

%\cite{Liang:2020dxr}
\bibitem{Liang:2020dxr}
W.~H.~Liang and E.~Oset,
%Observed $\Omega_b$ spectrum and meson-baryon molecular states,
Phys. Rev. D \textbf{101}, 054033 (2020)
%doi:10.1103/PhysRevD.101.054033
%[arXiv:2001.02929 [hep-ph]].
%8 citations counted in INSPIRE as of 10 Aug 2021

%%%%%%%%%%%%%%%%%%%%%%%%%%%%%Xic%%%%%%%%%%%%%%%%%%%%%%%%%%%%%%%%%%%%%

%\cite{Hu:2020zwc}
\bibitem{Hu:2020zwc}
X.~Hu, Y.~Tan and J.~Ping,
%Investigation of $\Xi_c^0$ in a chiral quark model,
Eur. Phys. J. C \textbf{81}, 370 (2021)
%doi:10.1140/epjc/s10052-021-09175-9
%[arXiv:2010.14984 [hep-ph]].
%0 citations counted in INSPIRE as of 10 Aug 2021

%\cite{Zhu:2020jke}
\bibitem{Zhu:2020jke}
H.~Zhu, N.~Ma and Y.~Huang,
%Description of the newly observed $\Xi_c^{0}$ states as molecular states,
Eur. Phys. J. C \textbf{80}, 1184 (2020)
%doi:10.1140/epjc/s10052-020-08747-5
%[arXiv:2005.02642 [hep-ph]].
%4 citations counted in INSPIRE as of 10 Aug 2021

%%%%%%%%%%%%%%%%%%%%%%%%%%%\Xi_b(6227)%%%%%%%%%%%%%%%%%%%%%%%%%%%%%%%%%%
%\cite{Huang:2018bed}
\bibitem{Huang:2018bed}
Y.~Huang, C.~j.~Xiao, L.~S.~Geng and J.~He,
%Strong decays of the $\Xi_b(6227)$ as a $\Sigma_b\bar{K}$ molecule,
Phys. Rev. D \textbf{99}, 014008 (2019)
%doi:10.1103/PhysRevD.99.014008
%[arXiv:1811.10769 [hep-ph]].
%18 citations counted in INSPIRE as of 10 Aug 2021

%\cite{Yu:2018yxl}
\bibitem{Yu:2018yxl}
Q.~X.~Yu, R.~Pavao, V.~R.~Debastiani and E.~Oset,
%Description of the $\Xi _c$ and $\Xi _b$ states as molecular states,
Eur. Phys. J. C \textbf{79}, 167 (2019)
%doi:10.1140/epjc/s10052-019-6665-z
%[arXiv:1811.11738 [hep-ph]].
%23 citations counted in INSPIRE as of 10 Aug 2021

%\cite{Wang:2020vwl}
\bibitem{Wang:2020vwl}
H.~J.~Wang, Z.~Y.~Di and Z.~G.~Wang,
%Analysis of the \ensuremath{\Xi}$_{b}$(6227) as the $\frac {1}{2}^{\pm }$ Pentaquark Molecular States with QCD Sum Rules,
Int. J. Theor. Phys. \textbf{59}, 3124 (2020)
%doi:10.1007/s10773-020-04566-2
%1 citations counted in INSPIRE as of 10 Aug 2021

%\cite{Zhu:2020lza}
\bibitem{Zhu:2020lza}
H.~Zhu and Y.~Huang,
%Radiative decay of $\Xi_b(6227)$ in a hadronic molecule picture,
Chin. Phys. C \textbf{44}, 083101 (2020)
%doi:10.1088/1674-1137/44/8/083101
%[arXiv:2004.00728 [hep-ph]].
%3 citations counted in INSPIRE as of 10 Aug 2021
%%%%%%%%%%%%%%%%%%%%%%%%%%%%%%%%%%%%%%%%%%%%%%%%%%%%%%%%%%%%%%%%%%%%%%%%%%%%%%

%\cite{Migura:2006ep}
\bibitem{Migura:2006ep}
  S.~Migura, D.~Merten, B.~Metsch and H.~R.~Petry,
%  Charmed baryons in a relativistic quark model,
  Eur.\ Phys.\ J.\ A {\bf 28}, 41 (2006)
%  doi:10.1140/epja/i2006-10017-9
%  [hep-ph/0602153].
  %%CITATION = doi:10.1140/epja/i2006-10017-9;%%
  %98 citations counted in INSPIRE as of 11 Jun 2021

%40\cite{Chen:2016iyi}
\bibitem{Chen:2016iyi}
  B.~Chen, K.~W.~Wei, X.~Liu and T.~Matsuki,
%  Low-lying charmed and charmed-strange baryon states,
  Eur. Phys. J. C  77, 154 (2017)
%  doi:10.1140/epjc/s10052-017-4708-x
%  [arXiv:1609.07967 [hep-ph]].
  %%CITATION = doi:10.1140/epjc/s10052-017-4708-x;%%
  %32 citations counted in INSPIRE as of 01 Apr 2020

%43\cite{Shah:2016mig}
\bibitem{Shah:2016mig}
  Z.~Shah, K.~Thakkar, A.~Kumar Rai and P.~C.~Vinodkumar,
%  Excited State Mass spectra of Singly Charmed Baryons,
  Eur.\ Phys.\ J.\ A {\bf 52},  313 (2016)
%  doi:10.1140/epja/i2016-16313-9
%  [arXiv:1602.06384 [hep-ph]].
  %%CITATION = doi:10.1140/epja/i2016-16313-9;%%
  %47 citations counted in INSPIRE as of 31 Jul 2020

%53\cite{Zhang:2008iz}
\bibitem{Zhang:2008iz}
  J.~R.~Zhang and M.~Q.~Huang,
%  Mass spectra of the heavy baryons $\Lambda_{Q}$ and $\Sigma_Q^{*}$ from QCD sum rules,
  Phys.\ Rev.\ D {\bf 77}, 094002 (2008)
%  doi:10.1103/PhysRevD.77.094002
%  [arXiv:0805.0479 [hep-ph]].
  %%CITATION = doi:10.1103/PhysRevD.77.094002;%%
  %26 citations counted in INSPIRE as of 10 Mar 2021

%%%%%%%%%%%%%%%%%%%%%%%%%%%%%%%%%%%%%%%%%%%%%%%%%%%%%%%%%%%%%%%%%%%%%%%%
%\cite{Tanabashi:2018oca}
\bibitem{Tanabashi:2018oca}
  P. A. Zyla {\it et al.} (Particle Data Group),
%  Review of Particle Physics,
  Prog.\ Theor. Exp. Phys. {\bf 2020}, 083C01 (2020)
  %doi:10.1103/PhysRevD.98.030001
  %%CITATION = doi:10.1103/PhysRevD.98.030001;%%
  %2401 citations counted in INSPIRE as of 06 Sep 2019

%31\cite{Mizuk:2004yu}
\bibitem{Mizuk:2004yu}
  R.~Mizuk {\it et al.} (Belle Collaboration),
%  Observation of an isotriplet of excited charmed baryons decaying to $\Lambda_c^+$ $\pi^-$,
  Phys.\ Rev.\ Lett.\  {\bf 94}, 122002 (2005)
%  doi:10.1103/PhysRevLett.94.122002
%  [hep-ex/0412069].
  %%CITATION = doi:10.1103/PhysRevLett.94.122002;%%
  %126 citations counted in INSPIRE as of 26 Jan 2021

%32\cite{Aubert:2008ax}
\bibitem{Aubert:2008ax}
  B.~Aubert {\it et al.} (BaBar Collaboration),
%  Measurements of $\mathcal{B}(\bar{B}^0 \to \Lambda_c^+ \bar{p})$ and $\mathcal{B}(B^- \to \Lambda_c^+ \bar{p} \pi^-)$ and Studies of $\Lambda_c^+$ $\pi^-$ Resonances,
  Phys.\ Rev.\ D {\bf 78}, 112003 (2008)
%  doi:10.1103/PhysRevD.78.112003
%  [arXiv:0807.4974 [hep-ex]].
  %%CITATION = doi:10.1103/PhysRevD.78.112003;%%
  %72 citations counted in INSPIRE as of 10 Feb 2021

%56\cite{Zhong:2007gp}
\bibitem{Zhong:2007gp}
  X.~H.~Zhong and Q.~Zhao,
%  Charmed baryon strong decays in a chiral quark model,
  Phys.\ Rev.\ D  77, 074008 (2008)
%  doi:10.1103/PhysRevD.77.074008
%  [arXiv:0711.4645 [hep-ph]].
  %%CITATION = doi:10.1103/PhysRevD.77.074008;%%
  %57 citations counted in INSPIRE as of 28 May 2018

 %33\cite{Chen:2007xf}
\bibitem{Chen:2007xf}
  C.~Chen, X.~L.~Chen, X.~Liu, W.~Z.~Deng and S.~L.~Zhu,
%  Strong decays of charmed baryons,
  Phys. Rev. D 75, 094017 (2007)
  %doi:10.1103/PhysRevD.75.094017
  %[arXiv:0704.0075 [hep-ph]].
  %%CITATION = doi:10.1103/PhysRevD.75.094017;%%
  %102 citations counted in INSPIRE as of 03 Apr 2020

%34\cite{Cheng:2006dk}
\bibitem{Cheng:2006dk}
  H.~Y.~Cheng and C.~K.~Chua,
%  Strong Decays of Charmed Baryons in Heavy Hadron Chiral Perturbation Theory,
  Phys. Rev. D  75, 014006 (2007)
 % doi:10.1103/PhysRevD.75.014006
 % [hep-ph/0610283].
  %%CITATION = doi:10.1103/PhysRevD.75.014006;%%
  %108 citations counted in INSPIRE as of 03 Apr 2020

%35\cite{Cheng:2015iom}
\bibitem{Cheng:2015iom}
  H.~Y.~Cheng,
%  Charmed baryons circa 2015,
  Front.\ Phys.\ (Beijing) {\bf 10}, 101406 (2015)
  %doi:10.1007/s11467-015-0483-z
  %%CITATION = doi:10.1007/s11467-015-0483-z;%%
  %41 citations counted in INSPIRE as of 08 Jan 2020

%37\cite{Gerasyuta:2007un}
\bibitem{Gerasyuta:2007un}
  S.~M.~Gerasyuta and E.~E.~Matskevich,
%  Charmed $(70,1^-)$ baryon multiplet,
  Int.\ J.\ Mod.\ Phys.\ E {\bf 17}, 585 (2008)
%  doi:10.1142/S0218301308010027
 % [arXiv:0709.0397 [hep-ph]].
  %%CITATION = doi:10.1142/S0218301308010027;%%
  %28 citations counted in INSPIRE as of 26 Jan 2021

%%%%%%%%%%%%%%%%%%%%%%%%%%%%%%%%%%%%%%%%%%%%%%%%%%%%%%%%%

%54\cite{GellMann:1962xb}
\bibitem{GellMann:1962xb}
  M.~Gell-Mann,
%  Symmetries of baryons and mesons,
  Phys. Rev. 125, 1067 (1962)
  %doi:10.1103/PhysRev.125.1067
  %%CITATION = doi:10.1103/PhysRev.125.1067;%%
  %1601 citations counted in INSPIRE as of 03 Apr 2020

%55\cite{Okubo:1961jc}
\bibitem{Okubo:1961jc}
  S.~Okubo,
%  Note on unitary symmetry in strong interactions,
  Prog. Theor. Phys.  27, 949 (1962)
  %doi:10.1143/PTP.27.949
  %%CITATION = doi:10.1143/PTP.27.949;%%
  %649 citations counted in INSPIRE as of 03 Apr 2020

%75\cite{Manohar:1983md}
\bibitem{Manohar:1983md}
  A.~Manohar and H.~Georgi,
%  Chiral Quarks and the Nonrelativistic Quark Model,
  Nucl.\ Phys.\ B {\bf 234}, 189 (1984)
%  doi:10.1016/0550-3213(84)90231-1
  %%CITATION = doi:10.1016/0550-3213(84)90231-1;%%
  %2129 citations counted in INSPIRE as of 10 Jun 2021

 %66\cite{Yao:2018jmc}
\bibitem{Yao:2018jmc}
  Y.~X.~Yao, K.~L.~Wang and X.~H.~Zhong,
%  Strong and radiative decays of the low-lying $D$-wave singly heavy baryons,
  Phys.\ Rev.\ D  98, 076015 (2018)
 % doi:10.1103/PhysRevD.98.076015
 % [arXiv:1803.00364 [hep-ph]].
  %%CITATION = doi:10.1103/PhysRevD.98.076015;%%
  %12 citations counted in INSPIRE as of 14 Jan 2020

%71\cite{Zhong:2008kd}
\bibitem{Zhong:2008kd}
  X.~H.~Zhong and Q.~Zhao,
%  Strong decays of heavy-light mesons in a chiral quark model,
  Phys.\ Rev.\ D  78, 014029 (2008)
  %doi:10.1103/PhysRevD.78.014029
  %[arXiv:0803.2102 [hep-ph]].
  %%CITATION = doi:10.1103/PhysRevD.78.014029;%%
  %75 citations counted in INSPIRE as of 26 May 2018
%%%%%%%%%%%%%%%%%%%%%%%%%%%%%%%%%%%%%%%%%%%%%%%%%%%%%%%%%%%%%%%%%%%%%%%%

%65\cite{Wang:2019uaj}
\bibitem{Wang:2019uaj}
  K.~L.~Wang, Q.~F.~L\"{u} and X.~H.~Zhong,
%  Interpretation of the newly observed $\Lambda_b(6146)^{0}$ and $\Lambda_b(6152)^0$ states in a chiral quark model,
  Phys.\ Rev.\ D 100, 114035 (2019)
  %doi:10.1103/PhysRevD.100.114035
  %[arXiv:1908.04622 [hep-ph]].
  %%CITATION = doi:10.1103/PhysRevD.100.114035;%%
  %2 citations counted in INSPIRE as of 14 Jan 2020

%67\cite{Xiao:2017udy}
\bibitem{Xiao:2017udy}
  L.~Y.~Xiao, K.~L.~Wang, Q.~F.~\"{u}, X.~H.~Zhong and S.~L.~Zhu,
%  Strong and radiative decays of the doubly charmed baryons,
  Phys.\ Rev.\ D 96, 094005 (2017)
  %doi:10.1103/PhysRevD.96.094005
  %[arXiv:1708.04384 [hep-ph]].
  %%CITATION = doi:10.1103/PhysRevD.96.094005;%%
  %39 citations counted in INSPIRE as of 14 Jan 2020

%68\cite{Xiao:2013xi}
\bibitem{Xiao:2013xi}
  L.~Y.~Xiao and X.~H.~Zhong,
%  $\Xi$ baryon strong decays in a chiral quark model,
  Phys.\ Rev.\ D  87, 094002 (2013)
  %doi:10.1103/PhysRevD.87.094002
  %[arXiv:1302.0079 [hep-ph]].
  %%CITATION = doi:10.1103/PhysRevD.87.094002;%%
  %23 citations counted in INSPIRE as of 14 Jan 2020

%69\cite{Liu:2019wdr}
\bibitem{Liu:2019wdr}
  M.~S.~Liu, K.~L.~Wang, Q.~F.~L\"{u} and X.~H.~Zhong,
%  $\Omega$ baryon spectrum and their decays in a constituent quark model,
  Phys.\ Rev.\ D {\bf 101}, 016002 (2020)
%  doi:10.1103/PhysRevD.101.016002
%  [arXiv:1910.10322 [hep-ph]].
  %%CITATION = doi:10.1103/PhysRevD.101.016002;%%
  %10 citations counted in INSPIRE as of 03 Mar 2021

%70\cite{Xiao:2018pwe}
\bibitem{Xiao:2018pwe}
  L.~Y.~Xiao and X.~H.~Zhong,
%  Possible interpretation of the newly observed $\Omega(2012)$ state,
  Phys.\ Rev.\ D {\bf 98}, 034004 (2018)
%  doi:10.1103/PhysRevD.98.034004
%  [arXiv:1805.11285 [hep-ph]].
  %%CITATION = doi:10.1103/PhysRevD.98.034004;%%
  %17 citations counted in INSPIRE as of 03 Mar 2021

%72\cite{Zhong:2009sk}
\bibitem{Zhong:2009sk}
  X.~H.~Zhong and Q.~Zhao,
%  Strong decays of newly observed $D_{sJ}$ states in a constituent quark model with effective Lagrangians,
  Phys.\ Rev.\ D 81, 014031 (2010)
  %doi:10.1103/PhysRevD.81.014031
  %[arXiv:0911.1856 [hep-ph]].
  %%CITATION = doi:10.1103/PhysRevD.81.014031;%%
  %52 citations counted in INSPIRE as of 14 Jan 2020

%73\cite{Zhong:2010vq}
\bibitem{Zhong:2010vq}
  X.~H.~Zhong,
%  Strong decays of the newly observed $D(2550)$, $D(2600)$, $D(2750)$, and $D(2760)$,
  Phys.\ Rev.\ D  82, 114014 (2010)
  %doi:10.1103/PhysRevD.82.114014
  %[arXiv:1009.0359 [hep-ph]].
  %%CITATION = doi:10.1103/PhysRevD.82.114014;%%
  %49 citations counted in INSPIRE as of 14 Jan 2020

%74\cite{Xiao:2014ura}
\bibitem{Xiao:2014ura}
  L.~Y.~Xiao and X.~H.~Zhong,
%  Strong decays of higher excited heavy-light mesons in a chiral quark model,
  Phys.\ Rev.\ D 90, 074029 (2014)
  %doi:10.1103/PhysRevD.90.074029
  %[arXiv:1407.7408 [hep-ph]].
  %%CITATION = doi:10.1103/PhysRevD.90.074029;%%
  %30 citations counted in INSPIRE as of 14 Jan 2020

%76\cite{Zhao:2002id}
\bibitem{Zhao:2002id}
  Q.~Zhao, J.~S.~Al-Khalili, Z.~P.~Li and R.~L.~Workman,
%  Pion photoproduction on the nucleon in the quark model,
  Phys.\ Rev.\ C {\bf 65}, 065204(2002)
%  doi:10.1103/PhysRevC.65.065204
%  [nucl-th/0202067].
  %%CITATION = doi:10.1103/PhysRevC.65.065204;%%
  %56 citations counted in INSPIRE as of 10 Jun 2021

%77\cite{Li:1994cy}
\bibitem{Li:1994cy}
  Z.~P.~Li,
%  The Threshold pion photoproduction of nucleons in the chiral quark model,
  Phys.\ Rev.\ D {\bf 50}, 5639 (1994)
%  doi:10.1103/PhysRevD.50.5639
%  [hep-ph/9404269].
  %%CITATION = doi:10.1103/PhysRevD.50.5639;%%
  %46 citations counted in INSPIRE as of 10 Jun 2021

%78\cite{Li:1997gd}
\bibitem{Li:1997gd}
  Z.~P.~Li, H.~X.~Ye and M.~H.~Lu,
%  An Unified approach to pseudoscalar meson photoproductions off nucleons in the quark model,
  Phys.\ Rev.\ C {\bf 56}, 1099 (1997)
%  doi:10.1103/PhysRevC.56.1099
%  [nucl-th/9706010].
  %%CITATION = doi:10.1103/PhysRevC.56.1099;%%
  %92 citations counted in INSPIRE as of 10 Jun 2021



\end{thebibliography}
\end{document}